\documentclass[%
reprint,
superscriptaddress,
 amsmath,amssymb,
 aps,
]{revtex4-2}

\usepackage[dvipdfmx]{graphicx}
\usepackage{dcolumn}
\usepackage{bm}
\usepackage{color}
\usepackage{hyperref}
\usepackage{physics}
\usepackage{orcidlink}

\newcommand{\mapsquared}{\langle\mathcal{M}_{\rm ap}^2\rangle}
\newcommand{\mapcube}{\langle\mathcal{M}_{\rm ap}^3\rangle}

\usepackage{afterpage}
\usepackage{lineno}

\begin{document}

\preprint{APS/123-QED}

\title{Tidal alignment and tidal torquing modeling for the cosmic shear three-point correlation function and mass aperture skewness}

\author{Rafael C. H. Gomes\orcidlink{0000-0002-3800-5662}}
\author{Kyle Miller}
\author{Sunao Sugiyama}
\affiliation{Department of Physics and Astronomy, University of Pennsylvania, Philadelphia, PA 19104, USA}
\author{Jonathan Blazek}
\affiliation{Department of Physics, Northeastern University, Boston, MA 02115, USA}
\author{Thomas Bakx}
\affiliation{Institute for Theoretical Physics, Utrecht University, Princetonplein 5, 3584 CC, Utrecht, The Netherlands}
\author{Bhuvnesh Jain}
\affiliation{Department of Physics and Astronomy, University of Pennsylvania, Philadelphia, PA 19104, USA}
\date{\today}

\begin{abstract}
We present a model for the intrinsic alignment contamination of the shear three-point correlation function and skewness of the mass aperture statistic using the tidal alignment and tidal torquing (TATT) formalism. We compute the intrinsic alignment bispectra components in terms of the TATT model parameters. We consider two effective field theory approaches in the literature, relate them to the TATT model parameters and an extension to TATT that includes the velocity-shear (VS) parameter. We compare the impact of changing between NLA, TATT, and TATT+VS on the theoretical computation of the 3PCF using the best fit parameters and tomographic redshift distributions from Dark Energy Survey Year 3. We find that the TATT model significantly impacts the skewed triangle configurations of the 3PCF. Additionally, including the higher-order effects from TATT can introduce opposite effects on the two-point function and on the mass aperture skewness, damping the signal of the former while boosting the signal of the latter. We argue that a joint 2PCF+3PCF analysis with the TATT model can help break the degeneracy between its model parameters and provide more robust constraints on both cosmology and intrinsic alignment amplitude parameters. We show that typical values of order  unity for the intrinsic alignment parameters introduce differences of around $10\%$ between NLA and TATT predictions. 
\end{abstract}

\maketitle

\section{Introduction}\label{sec:introduction}

The study of weak lensing in photometric galaxy surveys has been responsible for increasingly precise constraints on the cosmological parameters. Of particular interest is the $S_8$ parameter, which is close to the optimal combination of the matter density $\Omega_m$ and the amplitude of density fluctuations $\sigma_8$ for which weak lensing probes can put a tightest constraint. Results from the Dark Energy Survey (DES) \cite{Secco.Samuroff}, the Kilo-Degree Survey (KiDS) \cite{Asgari.KiDS}, and the Hyper-Suprime Camera (HSC) \cite{Dalal.HSC} have all recently placed constraints in $S_8$ at a precision level of $\approx 3\%$. These constraints, however, place themselves in a tension of the level of $2$-$3\sigma$ with Cosmic Microwave Background (CMB) constraints obtained through the Planck satellite (with the latest KiDS results being in closer agreement with the CMB, as mentioned below).

The identification of the $S_8$ tension highlights the importance of accurate modeling at all stages of a cosmological analysis, from redshift calibration to systematic error mitigation. It also motivates us to develop ways of further increasing the precision of our constraints. While changes in methodology can be responsible for a significant reduction of the tension, as found by \citet{KiDS.Legacy} with KiDS data, a definitive answer of whether the tension can be solely attributed, across all data sets, to methodological issues, still has to be investigated. 

Traditional analyses of cosmic shear data rely on two-point statistics, which capture the Gaussian features of the shear field. Several methods have been proposed to go beyond two-point information and access the non-Gaussian information produced by the non-linear development of density perturbations \cite{Petri_2013,Cheng.Yuan-Sen, Allys_2020, barthelemy2024makingleapimodelling, Zurcher.Fluri.2022, Giblin_2023, Marques_2024,Heydenreich_2021_homology,anbajagane20233rdmomentpracticalstudy, Gong_2023, Jeffrey.Whiteway.Gatti.2024, Prat.Gatti.Doux.2025, Gebauer.3PCF}. In this context, the three-point correlation function (3PCF) emerges as a natural higher-order counterpart of the two-point function, enabling a theoretical framework that facilitates their joint modeling. By compressing the information of the 3PCF into the skewness of the mass aperture statistic, cosmological analyses have been performed on DES \cite{Gomes.DES.methdos}\cite{Gomes.DES.data}, HSC \cite{Sugiyama.HSC.3PCF}, and KiDS \cite{Burger.Martinet.2023} data, yielding improvements of $111\%$, $80\%$ and $93\%$ on the joint constraint on $\Omega_m$ and $S_8$. 

Moving from Stage III to Stage IV surveys, the impact of higher-order statistics is expected to be significantly larger, with a forecast from the Euclid collaboration predicting an improvement of $251\%$ when using the mass aperture skewness \cite{Euclid.forecast}. With the high level of data expected from Euclid, LSST, and Roman, a robust joint modeling of galaxy intrinsic alignments becomes a crucial step to guarantee unbiased cosmological constraints. 

Two-point statistical analyses of Stage III datasets have mainly made use of two models for intrinsic alignments: the non-linear alignment (NLA), and the tidal alignment and tidal torquing (TATT) models. The former builds itself from the assumption that galaxy ellipticities are linearly related to the local gravitational potential, with the density perturbations described through the non-linear matter power spectrum. The latter allows for higher-order effects, including tidal torquing, being built from a perturbative expansion of the density and tidal fields. While both models typically assume a power-law for redshift evolution of intrinsic alignments, this assumption can also be replaced by a non-trivial relation, as done by \citet{Chen.deRose.Zhou.2024}.

For third-order shear statistics, current analyses have relied on the simpler NLA model, which can be naturally extended from the power spectrum to the bispectrum. If the alignment amplitude is high, however, this choice to neglect higher-order terms can induce substantial biases. This effect was studied at the level of two-point statistics by \citet{Secco.Samuroff}.

The modeling of the intrinsic alignment bispectrum involves the computation of several distinct components.
The density-density-shape component was studied by \citet{Schmitz_2018} following the TATT formalism with an extension to include velocity shear. An EFT modeling of the intrinsic alignment bispectrum was developed by \citet{Vlah_2020} and further studied by \citet{Vlah_2021}. More recently, \citet{bakx} provide the complete expressions for the anisotropic IA bispectrum with an implementation using FFTs and a forecast of high signal-to-noise detection on Stage IV spectroscopic surveys.

The use of an EFT modeling for two-point and three-point statistics yields a substantial number of nuisance parameters in the cosmological inference process. Choices such as NLA and TATT provide a more restricted parameter space which can be more desirable in order to avoid unwanted degeneracies between nuisance parameters and cosmological parameters. We are motivated to explore TATT at the level of the 3PCF in order to add some model complexity relative to NLA (two more amplitude parameters and one parameter for z-dependence). By introducing these parameters, we may be able not only to reduce biases on higher-order statistics constraints, but also to tighten the constraints on the IA parameters themselves via self-calibration. 

We are motivated to use TATT at the level of the 3PCF because of the different sensitivity of the intrinsic alignment parameters to the third-order and second-order shear information. This reveals the potential of this modeling not only to bring unbiased constraints from higher-order statistics but also to tighten the constraints on the IA parameters themselves.

In this paper, we review the different strategies for modeling the bispectra of intrinsic alignments, showing all the bispectra components in terms of the TATT parameters. We describe our implementation of the three-point correlation function and of the skewness of the mass aperture statistic of intrinsic alignments, providing a clear pathway towards application of TATT on joint second and third-order shear analyses. We show the level of IA contamination expected at the level of the 3PCF given parameter values estimated from the 2PCF on DES Year 3 data. 

\section{Modeling}

\subsection{Cosmic shear two-point and three-point statistics}

The characterization of weak lensing allows us to directly probe the dark matter density distribution in the region between the observed source galaxies and the observer. N-point statistics are modeled from the convergence field, which characterizes the isotropic magnification of the background galaxy's apparent size. We write the lensing efficiency at comoving distance $\chi$ for an $a$-th tomographic bin
\begin{equation}
    q_{a}(\chi) = \int_\chi^\infty \dd\chi' p_{a}(\chi')\frac{\chi'-\chi}{\chi'},
\end{equation}
where the redshift distribution $p_a(\chi)$ is normalized as $\int{\rm d}\chi p_a(\chi)=1$. The lensing convergence field is obtained by integrating the matter density field with the associated lensing efficiency kernel \cite{Kilbinger_2015}:
\begin{equation}
    \kappa_{a}(\bm{X}) = \frac{3\Omega_{\rm m}H_0^2}{2c^2}\int_0^{\infty}\dd\chi ~q_{a}(\chi)\frac{\delta_{\rm m}\left(\chi\bm{X}, \chi; z(\chi)\right)}{a(\chi)}.
\end{equation}
Here $z(\chi)$ is the redshift, $a(\chi)$ is the scale factor, and $\delta_{\rm m}(\bm{r}; z)$ is the matter density contrast at 3D coordinate $\bm{r}$ and redshift $z$. 

In Fourier space, we relate the convergence field to the shear field $\gamma(\ell)$ through the polar angle $\beta$ of the Fourier mode $\ell$:
\begin{align}
    \gamma_{\rm c}(\bm{\ell}) = \kappa(\bm{\ell})e^{2i\beta}.
\end{align}
Here the subscript c indicates that the shear field is defined in a Cartesian frame. The power spectrum of the convergence field is defined as 
\begin{equation}
    \langle\kappa_{a}(\bm{\ell}_1)\kappa_{b}(\bm{\ell}_2)\rangle 
    = (2\pi)^2\delta^{\rm D}(\bm{\ell}_1+\bm{\ell}_2)P_\kappa^{ab}(\ell_1),
    \label{eq:kappa-power}
\end{equation}
and is related to the matter power spectrum through integration with the lensing efficiency. Through the Limber approximation \cite{Kaiser_1992}, assuming tomographic redshift bins $a$ and $b$, we have
\begin{equation}
P_{\kappa}^{ab}(\ell) = \frac{9\Omega_{\rm m}^2H_0^4}{4c^4}\int_{0}^{\infty}\dd\chi\frac{q_a(\chi)q_b(\chi)}{a^2(\chi)}P_{\delta}\left(\frac{\ell}{\chi},z(\chi)\right).
\end{equation}

Analogously, for three-point statistics, we define the convergence bispectrum as
\begin{align}
\langle\kappa_a(\bm{\ell}_1)\kappa_b(\bm{\ell}_2)\kappa_c(\bm{\ell}_3)\rangle 
    =& (2\pi)^2\delta^{\rm D}\left(\bm{\ell}_1 + \bm{\ell}_2 + \bm{\ell}_3\right)\nonumber\\
    &\times B_\kappa^{abc}(\ell_1, \ell_2, \ell_3).
\end{align}
The convergence bispectrum is also related to the matter bispectrum. Correlating information across three redshift bins $a$, $b$, and $c$, under the Limber approximation, we write \cite{Buchalter_2000}
\begin{align}
\begin{split}
    B_\kappa^{abc}(\ell_1, \ell_2, \ell_3) &= \frac{27\Omega_{\rm m}^3H_0^6}{8c^6}\int_{0}^{\infty}\dd\chi\frac{q_a(\chi)q_b(\chi)q_c(\chi)}{a(\chi)^3\chi}\\
    &\hspace{2em}\times B_{\delta}\left(\frac{\ell_1}{\chi},\frac{\ell_2}{\chi},\frac{\ell_3}{\chi},z(\chi)\right).
\end{split}
\end{align}

In real space, the shear two-point and three-point correlation functions are functions of the power spectrum and bispectrum. The shear field is a spin-2 field, yielding three two-point functions: $\langle\gamma_t\gamma_t\rangle$, $\langle\gamma_t\gamma_{\cross}\rangle$, and $\langle\gamma_{\cross}\gamma_{\cross}\rangle$. Of these, we have only two independent components, since in a universe with parity symmetry the correlation between the tangential and radial components is null \cite{Kilbinger_2015}. The functions $\xi_{\pm}$ are commonly used for cosmological inference and are given by
\begin{align}
    \xi_{+}(\theta) \equiv \langle\gamma_t\gamma_t\rangle + \langle\gamma_{\cross}\gamma_{\cross}\rangle , \\
    \xi_{-}(\theta) \equiv \langle\gamma_t\gamma_t\rangle - \langle\gamma_{\cross}\gamma_{\cross}\rangle.
\end{align}
Here it is convenient to separate the convergence field into a curl-free and a gradient-free component, introducing E and B modes. The two-point correlation functions can be written in terms of the E and B modes of the convergence power spectrum \cite{Kilbinger_2015}:
\begin{align}
    \xi_{+}(\theta) = \int_0^{\infty} \frac{\ell d\ell}{2\pi} J_0(\ell\theta)[P_{\kappa}^E(\ell)+P_{\kappa}^B(\ell)],\\
    \xi_{-}(\theta) = \int_0^{\infty} \frac{\ell d\ell}{2\pi} J_4(\ell\theta)[P_{\kappa}^E(\ell)-P_{\kappa}^B(\ell)],
\end{align}
where $J_{0/4}(x)$ is the 0th-/4th-order Bessel function of the first kind, and the auto E/B mode power spectrum, $P_\kappa^{E/B}$, is defined similarly as Eq~(\ref{eq:kappa-power}) but with E/B modes of convergence field.

To model three-point statistics, we can write correlations between the tangential and radial components of the shear at three distinct points. Following \citet{Schneider.Lombardi.2002}, we define the natural components of cosmic shear as

\begin{equation}
\begin{split}
\Gamma^0 = \langle\gamma(\textbf{X}_1)\gamma(\textbf{X}_2)\gamma(\textbf{X}_3)\rangle, \\
\Gamma^1 = \langle\gamma^*(\textbf{X}_1)\gamma(\textbf{X}_2)\gamma(\textbf{X}_3)\rangle, \\
\Gamma^2 = \langle\gamma(\textbf{X}_1)\gamma^*(\textbf{X}_2)\gamma(\textbf{X}_3)\rangle, \\
\Gamma^3 = \langle\gamma(\textbf{X}_1)\gamma(\textbf{X}_2)\gamma^*(\textbf{X}_3)\rangle.
\end{split}
\end{equation}

In terms of the radial and tangential components, we have
\begin{align}
&\begin{cases}
\textbf{Re}(\Gamma^0) = \gamma_{ttt} - \gamma_{t\times\times} - \gamma_{\times t \times} - \gamma_{\times\times t} \\
\textbf{Im}(\Gamma^0) = \gamma_{tt\times} + \gamma_{t\times t} + \gamma_{\times t t} - \gamma_{\times\times \times}
\end{cases}\\
&\begin{cases}
\textbf{Re}(\Gamma^1) = \gamma_{ttt} - \gamma_{t\times\times} + \gamma_{\times t \times} + \gamma_{\times\times t} \\
\textbf{Im}(\Gamma^1) = \gamma_{tt\times} + \gamma_{t\times t} - \gamma_{\times t t} + \gamma_{\times\times \times}
\end{cases}\\
&\begin{cases}
\textbf{Re}(\Gamma^2) = \gamma_{ttt} + \gamma_{t\times\times} -\gamma_{\times t \times} + \gamma_{\times\times t} \\
\textbf{Im}(\Gamma^2) = \gamma_{tt\times} - \gamma_{t\times t} + \gamma_{\times t t} + \gamma_{\times\times \times}
\end{cases}\\
&\begin{cases}
\textbf{Re}(\Gamma^3) = \gamma_{ttt} + \gamma_{t\times\times} +\gamma_{\times t \times} - \gamma_{\times\times t} \\
\textbf{Im}(\Gamma^3) = - \gamma_{tt\times} + \gamma_{t\times t} + \gamma_{\times t t} + \gamma_{\times\times \times}.
\end{cases}
\end{align}
The functions $\Gamma^0$, $\Gamma^1$, $\Gamma^2$ and $\Gamma^3$ can be worked out as functions of the convergence bispectrum [see \cite{Schneider.Kilbinger.2005}\cite{ Heydenreich.Schneider.2022} for detailed derivation].
A fast computational method for the necessary highly oscillatory integrals is proposed by \citet{sugiyama2024fastmodelingshearthreepoint} using a multipole expansion of the bispectrum. 

Finally, cosmological analyses with the three-point correlation function require significant data compression. The skewness of the mass aperture statistic ($\mapcube$) \cite{Jarvis.Jain.2003} is a physically motivated way of compressing the full information content of the three-point correlation function. It also ensures separation of E and B modes. While the lensing signal by itself is not expected to have B modes, this is not the case for intrinsic alignments. Such a statistic is, therefore, optimal to be used in conjunction with complex alignment models in which there may be non-zero presence of B modes.

We follow \citet{Jarvis.Jain.2003} and define the mass aperture $M_{\text{ap}}$ in terms of the shear field $\gamma$ as
\begin{equation}
M_{\text{ap}}(R)\int \text{d}^2rQ_R(r)\gamma_t(\boldsymbol{r}),
\end{equation}
where
\begin{equation}
Q_R(r)=\frac{r^2}{4\pi R^4}\exp\left(\frac{-r^2}{2R^2}\right).
\end{equation}
In this way, the mass aperture represents a single measurement of the convergence signal within a circular patch. The second-order and third-order shear information is found by taking the variance ($\mapsquared$) and the skewness ($\mapcube$) of the mass aperture. Measuring the skewness directly from convergence maps is not feasible for realistic survey data because it introduces the necessity of accounting for the complicated survey geometry at the level of theoretical modeling. To circumvent this issue, one can compute $\mapcube$ as a function of the natural components of the three-point correlation function, which can be measured from survey data regardless of the survey edges and holes. The theoretical model for $\mapcube$ is thus performed consistently, from the binned predictions of the full 3PCF. The efficiency of $\mapcube$ to compress the 3PCF data was demonstrated in a principal component analysis by \citet{Heydenreich.Schneider.2022}.

\subsection{Review of intrinsic alignment bispectra models}

Galaxy intrinsic alignments appear as contaminants to weak lensing measurements. Shear estimators are based on ellipticity measurements, and therefore do not separate the actual shape distortion due to weak gravitational lensing from the correlated intrinsic shapes of the galaxies. We can split the measured signal into its different contributions and model them separately. The relation between the lensing-induced ellipticity and shear is given by the response matrix as $\epsilon = R\gamma$. The intrinsic alignment contamination is additive at the level of the ellipticity. We perform our model, however, at the level of the shear $\gamma$, absorbing the difference between $\epsilon$ and $\gamma$ into the definition of the intrinsic alignment model parameters.

Thus, at the level of the galaxy shear, we model $\gamma$ as:
\begin{align}
\gamma = \gamma_{\rm G} + \gamma_I
\end{align}
where the first term is the gravitational lensing shear and the second is the intrinsic galaxy shape alignment to the underlying tidal field. 

In the context of photometric surveys, the correlations will be measured at a set of tomographic redshift bins. Correlations between intrinsic ellipticities at one bin and shear at another can be non-zero due to the same dark matter overdensity inducing alignment locally and shear on a higher redshift bin. For two-point statistics, the ellipticity correlation between tomographic bins $a$ and $b$ is
\begin{equation}
\langle \gamma^a \gamma^b \rangle = \langle \gamma_{\rm G}^a \gamma_{\rm I}^b \rangle + \langle \gamma_{\rm G}^a \gamma_{\rm G}^b \rangle + \langle \gamma_{\rm I}^a \gamma_{\rm G}^b \rangle + \langle \gamma_{\rm I}^a \gamma_{\rm I}^b \rangle
\end{equation}
where $\gamma$ stands for the cosmic shear, and $I$ for the intrinsic shape correlations \cite{Lamman_2024}. 

Similarly, for three-point statistics, we have
\begin{align}
\langle \gamma^a \gamma^b \gamma^c\rangle 
 =&\langle \gamma_{\rm G}^a \gamma_{\rm G}^b \gamma_{\rm G}^c\rangle \nonumber\\
&+ \langle \gamma_{\rm I}^a \gamma_{\rm G}^b \gamma_{\rm G}^c\rangle 
 + \langle \gamma_{\rm G}^a \gamma_{\rm I}^b \gamma_{\rm G}^c\rangle 
 + \langle \gamma_{\rm G}^a \gamma_{\rm G}^b \gamma_{\rm I}^c\rangle \nonumber\\
&+ \langle \gamma_{\rm I}^a \gamma_{\rm I}^b \gamma_{\rm G}^c\rangle 
 + \langle \gamma_{\rm I}^a \gamma_{\rm G}^b \gamma_{\rm I}^c\rangle 
 + \langle \gamma_{\rm G}^a \gamma_{\rm I}^b \gamma_{\rm I}^c\rangle \nonumber\\
&+ \langle \gamma_{\rm I}^a \gamma_{\rm I}^b \gamma_{\rm I}^c\rangle 
\label{threepointshear}
\end{align}
The IA contamination to the cosmic shear signal can be modeled by computing the correlation functions involving the intrinsic shapes of galaxies. The different components of Eq.~\ref{threepointshear} then translate into corresponding additive contributions to the total contamination signal of the convergence bispectrum.

The non-linear alignment model, NLA in short,  is based on the assumption that galaxy ellipticities are linearly related to the trace-free part of the second derivative of the local gravitational potential, i.e. the tidal field, with the density perturbation described through the non-linear matter power spectrum \cite{Hirata04,Bridle_2007}. Typically, the proportionality factor $f_{\text{IA}}$ is written in terms of an amplitude parameter $A_{\text{IA}}$ and a second parameter $\alpha_{\text{IA}}$ encoding redshift dependence as

\begin{equation}
f_{\text{IA}}(z) = - A_{\text{IA}}\left(\frac{1+z}{1+z_0}\right)^{\alpha_{\text{IA}}}\frac{\bar{C_1}\Omega_m\rho_{\text{crit}}}{D(z)}
\label{nlaamplitude}
\end{equation}
where $D(z)$ is the linear growth factor, $\rho_{\text{crit}}$ is the critical density, and $\bar{C_1}$ is a normalization factor, usually set to $5 \times 10^{-14} \text{Mpc}^3/(h^2M_{\odot})$ \cite{Gong_2023}.

In the NLA model, because of this proportionality of the IA term to the non-linear matter density field at each redshift, to which the lensing convergence field is also proportional, 
the observed galaxy shear can be modeled just by replacing the lensing kernel with an additional kernel due to intrinsic alignment as
\begin{align}
    q_a(\chi) \rightarrow q_a(\chi) + f_{\rm IA}(z) p_a(\chi)\frac{{\rm d}z}{{\rm d}\chi}
    \label{nlakernel}
\end{align}
where the Jacobian ${\rm d}z/{\rm d}\chi$ is needed in order to change the integral variable from the redshift $z$ to the comoving distance $\chi$ for intrinsic alignment term.
The modeling advantage of this approach is that the actual matter bispectrum prescription does not need to be modified. However, it falls short of a complete characterization of the galaxy response to the 3D tidal field, not allowing for effects such as tidal torquing. 

The non-linear alignment approach for the bispectrum was used on DES Y3 data for the integrated three-point correlation function analysis \cite{Gebauer.3PCF} and for the skewness of the mass aperture statistic analysis \cite{Gomes.DES.data}. It was also used for the $\mapcube$ analyses of KiDS \cite{Burger.Martinet.2023} and HSC \cite{Sugiyama.HSC.3PCF}.

A more complex approach to intrinsic alignments is performed in the TATT model, which was introduced by \citet{Blazek.2019} and performs a characterization of the tidal field dependence of the correlated intrinsic shapes. The 3D intrinsic shape $\gamma^I_{ij}$ is perturbatively expanded and written in terms of the tidal tensor $s_{ij}$. The terms included in the TATT model are a linear term $C_1$, a quadratic term $C_2$, and a density weighting term $C_{1\delta}$, which accounts for the effect that the positions where we can detect galaxy shapes are only those positions where galaxies exist. We have:
\begin{equation}
\gamma^I_{ij} = C_1 s_{ij} + C_{1\delta}(\delta s_{ij}) + C_2\left(s_{ik}s_{kj}-\frac{1}{3}\delta_{ij}s^2\right).
\label{tatt_def}
\end{equation}
In this formalism, the $C_1$ parameter can be written in terms of an amplitude parameter and  a redshift evolution parameter, following Eq.~(\ref{nlaamplitude}), in which $C_1$ is identified with the $f_{\text{IA}}$ parameter. The TATT implementation used in the DES Y3 analysis of \cite{Secco.Samuroff} considers $C_{1\delta}$ to be related to $C_1$ via the free linear bias parameter $b_{\text{TA}}$ (e.g.\ \cite{Blazek.2015} as
\begin{equation}
C_{1\delta} = b_{\text{TA}}C_1.
\end{equation}
Finally, the quadratic term can also be modeled via a fixed redshift evolution parameter $\alpha_2$ as
\begin{equation}
C_2 = 5A_2\left(\frac{1+z}{1+z_0}\right)^{\alpha_{2}}\frac{\bar{C_1}\Omega_m\rho_{\text{crit}}}{D(z)^2}.
\end{equation}
To summarize, $(A_1, A_2, b_{\rm TA}, \alpha_1, \alpha_2)$ is the widely-used set of TATT model parameters in the literature, although other treatments of redshift dependence of each term could be consistently applied.

The first step towards the development of TATT for the bispectrum was taken by \citet{Schmitz_2018}. As we describe below, they also considered the velocity-shear effect mediated by the  $t_{ij}$ tensor [see Eq.~\ref{tatt_vs_def}]. They used standard perturbation theory (SPT) to compute a tree-level density-density-shape bispectrum (the $B_{ggI}$ component). More general expressions were developed within the effective field theory (EFT) framework, of which both TATT and NLA can be interpreted as subsets, given the caveat that, unlike EFT, they phenomenologically extend the model to smaller scales by replacing the linear matter power spectrum $P(k)$ by the non-linear $P_{NL}(k)$. A similar extension at the level of the bispectrum is proposed in Section~\ref{sec:tatt-ia-bispectrum}. In \citet{Vlah_2020}, the bias expansion for the three-dimensional galaxy shapes is written down, and expressions are devised for 3D two-point correlations (at one-loop) and for three-point correlation functions (at tree level). At tree-level, the EFT approach for the 3PCF is complete up to the second order of the field expansion, also accounting for the stochasticity of the shape perturbations. The velocity shear extension to TATT (TATT+VS) is also complete to second order in the field, although it lacks the stochastic terms.

A complete EFT modeling of the anisotropic IA bispectrum is presented by \citet{bakx} and validated for large scales with N-body simulations from the \texttt{DarkQuest} project \cite{Nishimichi_2019} by \citet{bakx2025_2}. The shared IA parameters between two-point and three-point correlations are found to be consistent with one another.

The EFT parametrization starts by defining the shape tensor $S_{ij}(\boldsymbol{x})$ by
\begin{equation}
S_{ij}(\textbf{x}) = \frac{1}{3}\delta_{ij}\delta_S(\textbf{x}) + g_{ij}(\textbf{x}),
\end{equation}
where $\delta_S$ and $g_{ij}$ are the matter density perturbation and the intrinsic shape perturbation. The complete bispectrum is computed as the expectation value of the Fourier transform of the shape tensor, which we denote as $\tilde{S}(\textbf{k})$. Thus, we have
\begin{equation}
\begin{split}
(2 \pi)^3 \delta^{\rm D}(\boldsymbol{k}_1+\boldsymbol{k}_2+\boldsymbol{k}_3) B_{ijklrs}(\boldsymbol{k}_1,\boldsymbol{k}_2,\boldsymbol{k}_3) 
 \\ = 
\langle \tilde{S}_{ij}{(\boldsymbol{k}_1)\tilde{S}_{kl}(\boldsymbol{k}_2)\tilde{S}_{rs}(\boldsymbol{k}_3)}\rangle
\end{split}
\end{equation}
By replacing the full shape tensors inside the expectation value with combinations of the scalar perturbation $\delta$ and the tensor perturbation $g_{ij}$, four different bispectrum components arise, which are denoted by \citet{bakx} as $B^{sss}$, $B^{ssg}$, $B^{sgg}$, and $B^{ggg}$, where the $s,g$ indices refer respectively to scalar and tensor perturbations. The $B^{sss}$ gives rise to the usual cosmic shear bispectrum, as under standard assumptions it is equivalent to the scalar convergence bispectrum, and the other combinations are associated with the intrinsic alignment contamination terms from Eq. \ref{threepointshear}.

The bispectra can be projected to retain only the observable contributions. This gives origin to the separate bispectra for combinations of E, B, and scalar modes. The projection is defined by
\begin{equation}
\begin{split}
B^{XYZ}_{ijklrs}(\boldsymbol{k}_1,\boldsymbol{k}_2,\boldsymbol{k}_3) 
&=\boldsymbol{M}^X_{ij}(\boldsymbol{k}_1)\boldsymbol{M}^Y_{kl}(\boldsymbol{k}_2)\boldsymbol{M}^Z_{rs}(\boldsymbol{k}_3) \\
&\hspace{2em}\times B_{ijklrs}(\boldsymbol{k}_1,\boldsymbol{k}_2,\boldsymbol{k}_3)
\end{split}
\end{equation}
The projection operators are
\begin{equation}
\boldsymbol{M}^S_{ij}(\boldsymbol{\hat{k}}) = \delta_{ij},
\end{equation}
\begin{equation}
\begin{split}
\boldsymbol{M}^E_{ij}(\boldsymbol{\hat{k},\hat{n}}) = \frac{1}{2}(\boldsymbol{m}_i^-\boldsymbol{m}_j^-)\exp(-2i\phi_{\boldsymbol{k}})
\\ + \frac{1}{2}(\boldsymbol{m}_i^+\boldsymbol{m}_j^+)\exp(2i\phi_{\boldsymbol{k}}),
\end{split}
\end{equation}
and
\begin{equation}
\begin{split}
\boldsymbol{M}^B_{ij}(\boldsymbol{\hat{k},\hat{n}}) = \frac{1}{2i}(\boldsymbol{m}_i^-\boldsymbol{m}_j^-)\exp(-2i\phi_{\boldsymbol{k}})
\\ - \frac{1}{2i}(\boldsymbol{m}_i^+\boldsymbol{m}_j^+)\exp(2i\phi_{\boldsymbol{k}}),
\end{split}
\end{equation}
with the lower-case $\boldsymbol{m}(\boldsymbol{\hat{n}})$ operators given by $\boldsymbol{m}^- = (\boldsymbol{\hat{x}}+i\boldsymbol{\hat{y}})/\sqrt{2}$ and $\boldsymbol{m}^+ = (-\boldsymbol{\hat{x}}-i\boldsymbol{\hat{y}})/\sqrt{2}$.

The 3D bispectrum $B^{\alpha\beta\gamma}_{ijklrs}$, with the $\alpha,\beta,\gamma$ assuming each the value of either $s$ or $g$, will have deterministic and stochastic contributions. The deterministic contribution is given by
\begin{equation}
\begin{split}
&B^{\alpha\beta\gamma}_{ijklrs}(\boldsymbol{k}_1,\boldsymbol{k}_2, \boldsymbol{k}_3)\\
=&
2\mathcal{K}^{\alpha,(1)}_{ij}(\boldsymbol{k}_1)\mathcal{K}^{\beta,(1)}_{kl}(\boldsymbol{k}_2)\mathcal{K}^{\gamma,(2)}_{rs}(\boldsymbol{k}_1,\boldsymbol{k}_2)
P(k_1)P(k_2)\\
&+ \text{2 permutations}
\end{split}
\label{anisobisp}
\end{equation}
The kernel operators $\mathcal{K}$ depend on the $s,g$ indices. For scalar indices, we start with the first-order kernel:
\begin{equation}
    \mathcal{K}^{s,(1)}_{ij}(\boldsymbol{k}) = \frac{1}{3}\delta_{ij}b^s_1.
\end{equation}
The second-order kernel is:
\begin{equation}
    \mathcal{K}^{s,(2)}_{ij}(\boldsymbol{k}_1,\boldsymbol{k}_2) = \frac{1}{3}\delta_{ij}\left(b^s_1 F_2(\boldsymbol{k}_1,\boldsymbol{k}_2)+b^s_{2,1}\frac{(\boldsymbol{k}_1 \cdot\boldsymbol{k}_2)^2}{k_1^2 k_2^2}\right),
\end{equation}
where
\begin{equation}
    F_2(\bm{k}_1, \bm{k}_2) = \frac{5}{7} + \frac{1}{2}\hat{\bm{k}}_1\cdot\hat{\bm{k}}_2\left(\frac{k_1}{k_2} + 
    \frac{k_2}{k_1}\right) + \frac{2}{7}\hat{\bm{k}}_1\cdot\hat{\bm{k}}_2.
\end{equation}
For the tensor $g$ indices, the first order kernel is
\begin{equation}
    \mathcal{K}^{g,(1)}_{ij}(\boldsymbol{k}) = \left(\frac{\boldsymbol{k}_i\boldsymbol{k}_j}{k^2}-\frac{1}{3}\delta_{ij}\right)b^g_1.
\end{equation}
The second-order $\mathcal{K}^{g,(2)}_{ij}$ is achieved by taking the trace-free
component of the full $\mathcal{K'}^{g,(2)}_{ij}$ kernel, which is defined as
\begin{equation}
\begin{split}
    &\mathcal{K'}^{g,(2)}_{ij}(\boldsymbol{k}_1,\boldsymbol{k}_2) \\
    =& \frac{\boldsymbol{k}_{12,i}\boldsymbol{k}_{12,j}}{\boldsymbol{k}_{12}^2}F_2(\boldsymbol{k}_1,\boldsymbol{k}_2)(c^g_1+c^g_{2,1}) \\ 
    &+\frac{\boldsymbol{k}_1 \cdot \boldsymbol{k}_2 }{2k_1^2k_2^2}\left[(\boldsymbol{k}_{2,i}\boldsymbol{k}_{2,j} - \boldsymbol{k}_{1,i}\boldsymbol{k}_{1,j})c^g_{2,1}\right.\\
    &\hspace{5em} \left. +(\boldsymbol{k}_{1,i}\boldsymbol{k}_{2,j} + \boldsymbol{k}_{2,i}\boldsymbol{k}_{1,j})c^g_{2,2}\right] \\ 
    &+ \left(\frac{\boldsymbol{k}_{1,i}\boldsymbol{k}_{1,j}}{k_1^2}+\frac{\boldsymbol{k}_{2,i}\boldsymbol{k}_{2,j}}{k_2^2}\right)c^g_{2,3}
\end{split}
\end{equation}
The EFT expressions for the $B^{EEE}$, $B^{\delta EEE}$, and $B^{\delta \delta E}$ contributions can be computed under this formalism. To obtain reduced TATT expressions, we do not use the stochastic components of the EFT model. We also set $c^g_{2,1}=0$ because this term solely corresponds to the velocity shear effect. The remaining parameters can be transformed into those typically used in the context of TATT, as will be discussed in Section~\ref{sec:tatt-ia-bispectrum}. 

An alternative formalism for the EFT IA bispectrum was developed by \citet{Vlah_2021} for a tomographic projection, which is necessary in the context of photometric surveys. While the galaxy shapes are first projected onto the 2D observable sky with the E-B decomposition, the bispectrum at this point is still a function of three dimensional vectors. For a tomographic projection of this bispectrum at tree level, under the Limber approximation, we require integration with both the density and shape kernels. For the E-mode case, the expressions are 
\begin{widetext}
\begin{align}
B^{\delta \delta E}_{\text{proj}}(\boldsymbol{l}_1, \boldsymbol{l}_2, \boldsymbol{l}_3) 
& = \frac{1}{2} \times \int d\chi \frac{W_{\delta}(\chi)^2 W_g(\chi) }{\chi^4}(N_0 B_{002}^{(0)}-N_2 B_{002}^{(2)})(\boldsymbol{l}_1/\chi, \boldsymbol{l}_2/\chi, \boldsymbol{l}_3/\chi)
\\
B^{\delta EE}_{\text{proj}}(\boldsymbol{l}_1, \boldsymbol{l}_2, \boldsymbol{l}_3) 
& = \frac{N_0}{4} \times \int d\chi \frac{W_{\delta}(\chi)W_g(\chi)^2}{\chi^4}(N_0B_{022}^{(0,0)}-N_2 B_{022}^{(\{0,2\})})(\boldsymbol{l}_1/\chi, \boldsymbol{l}_2/\chi, \boldsymbol{l}_3/\chi)
\\
B^{EEE}_{\text{proj}}(\boldsymbol{l}_1, \boldsymbol{l}_2, \boldsymbol{l}_3) 
& = \frac{N_0^2}{8} \times \int d\chi \frac{W_g(\chi)^3}{\chi^4}(N_0B_{222}^{(0,0,0)}-N_2 B_{222}^{(\{0,0,2\})})(\boldsymbol{l}_1/\chi, \boldsymbol{l}_2/\chi, \boldsymbol{l}_3/\chi)
\end{align}
\end{widetext}
where $W_{\delta}$ and $W_g$ are the density and shape window functions, $N_0=\sqrt{3/2}$, $N_1=\sqrt{1/2}$, and $N_2=1$ are normalization constants, and the bispectrum terms $B_{\alpha\beta\gamma}$ are functions of the kernels $\mathcal{F}_0^{(0)}$ and $\mathcal{F}_2^{(m)}$. Here we note that our third equation differs from that of \citet{Vlah_2020} in that it replaces their $W_{\delta}W_g^2$ product with $W_g^3$. A contribution from the shape-shape-shape bispectrum should have factors of the shape window function alone. Our second equation also corrects the sign of their Eq.~4.18.

We now write the expression for each of the bispectrum terms. Here, we choose a convention for $\boldsymbol{k}_i$ where $\boldsymbol{k}_3$ is aligned with the x-axis and the orientation of $\boldsymbol{k}_1$, $\boldsymbol{k}_2$ and $\boldsymbol{k}_3$ is fixed as anticlockwise. For $B_{002}^{(m)}$, we have
\begin{equation}
\begin{split}
B_{002}^{\alpha\beta\gamma, (m)}(\boldsymbol{k}_1,\boldsymbol{k}_2,\boldsymbol{k}_3) &= \mathcal{F}_2^{\alpha\beta\gamma, (m)}(\boldsymbol{k}_1,\boldsymbol{k}_2)\\
&+\tilde{\delta}^K_{0,m}\mathcal{F}_0^{\beta\gamma\alpha, (0)}(\boldsymbol{k}_2,\boldsymbol{k}_3)\\ 
&+\tilde{\delta}^K_{0,m}\mathcal{F}_0^{\gamma\alpha\beta, (0)}(\boldsymbol{k}_3,\boldsymbol{k}_1),
\end{split}
\end{equation}
with $\tilde{\delta}^K_{0,m}=N_0^{-1}\delta^K_{0,m}$.

The remaining terms are
\begin{equation}
\begin{split}
B_{022}^{\alpha\beta\gamma, (m_2,m_3)}(\boldsymbol{k}_1,\boldsymbol{k}_2,\boldsymbol{k}_3)
&=\tilde{\delta}^K_{0,m_2}\mathcal{F}_2^{\beta\gamma\alpha, (m_3)}(\boldsymbol{k}_1,\boldsymbol{k}_2)\\
&+\tilde{\delta}^K_{0,m_2}\tilde{\delta}^K_{0,m_3}\mathcal{F}_0^{\alpha\beta\gamma, (0)}(\boldsymbol{k}_2,\boldsymbol{k}_3)\\
&+\tilde{\delta}^K_{0,m_3}\mathcal{F}_2^{\gamma\alpha\beta, (m_2)}(\boldsymbol{k}_3,\boldsymbol{k}_1),
\end{split}
\end{equation}
and
\begin{equation}
\begin{split}
&B_{222}^{\alpha\beta\gamma, (m_1,m_2,m_3)}(\boldsymbol{k}_1,\boldsymbol{k}_2,\boldsymbol{k}_3)\\
&= \tilde{\delta}^K_{0,m_1}\tilde{\delta}^K_{0,m_2}\mathcal{F}_2^{\alpha\beta\gamma, (m_3)}(\boldsymbol{k}_1,\boldsymbol{k}_2)\\
&+\tilde{\delta}^K_{0,m_2}\tilde{\delta}^K_{0,m_3}\mathcal{F}_2^{\beta\gamma\alpha, (m_1)}(\boldsymbol{k}_2,\boldsymbol{k}_3)\\
&+\tilde{\delta}^K_{0,m_3} \tilde{\delta}^K_{0,m_1}\mathcal{F}_2^{\gamma\alpha\beta, (m_2)}(\boldsymbol{k}_3,\boldsymbol{k}_1),
\end{split}
\label{b222vlah}
\end{equation}
where in Eq.~\ref{b222vlah} we reposition some of the indices that were misplaced in Eq.~5.35 of \citet{Vlah_2020}. Our updated version of these equations makes this formalism consistent with that of \citet{bakx}.

When dealing only with correlations between shape and matter density, the expressions for the $\mathcal{F}$ kernels can be written in a simplified manner. To do this, we do not include the galaxy bias expansion of \citet{Vlah_2020} but instead, set their $b_1^s$ parameter to unity and ignore the higher-order $b^s_{2,1}$ and $b^s_{2,2}$ terms. We have
\begin{equation}
\mathcal{F}_0^{\alpha\beta\gamma,(0)}(\boldsymbol{k}_1,\boldsymbol{k}_2) = 2F_2(\boldsymbol{k}_1,\boldsymbol{k}_2)P_L(k_1)P_L(k_2),
\end{equation}
\begin{equation}
\begin{split}
&\mathcal{F}_2^{\alpha\beta\gamma,(0)}(\boldsymbol{k}_1,\boldsymbol{k}_2) = \frac{4N_0}{3}P_L(k_1)P_L(k_2) \\ &\times \left\{\vphantom{\frac{DONT}{USE}} c_1^g F_2(\boldsymbol{k}_1,\boldsymbol{k}_2) 
+c_{2,1}^g \bigg(\frac{5}{7}(1-(\hat{\boldsymbol{k}}_1 \cdot \hat{\boldsymbol{k}}_2)^2) \right.  \\ 
&\hspace{12em}+\frac{1}{4}(\hat{k}_{1x}^2-\hat{k}_{1y}^2+\hat{k}_{2x}^2-\hat{k}_{2y}^2)\\  &\hspace{12em}+\frac{1}{2}((\hat{\boldsymbol{k}}_1 \cdot \hat{\boldsymbol{k}}_2)\hat{k}_{1x}\hat{k}_{2x})\bigg) \\
&\hspace{2em}+c_{2,2}^g\bigg(\frac{1}{4}(\hat{k}_{1x}^2-\hat{k}_{1y}^2+\hat{k}_{2x}^2-\hat{k}_{2y}^2)\\&\hspace{7em}+\frac{1}{2}((\hat{\boldsymbol{k}}_1 \cdot \hat{\boldsymbol{k}}_2)\hat{k}_{1x}\hat{k}_{2x})\bigg) \\
&\hspace{2em}+\left. c_{2,3}^g\left(\frac{1}{4}(2\hat{k}_{1x}^2-\hat{k}_{1y}^2+2\hat{k}_{2x}^2-\hat{k}_{2y}^2)\right)\right\},
\end{split}
\end{equation}
\begin{equation}
\begin{split}
&\mathcal{F}_2^{\alpha\beta\gamma,(1)}(\boldsymbol{k}_1,\boldsymbol{k}_2) = -\sqrt{2}N_1 P_L(k_1)P_L(k_2) \\ &\times (c_{2,1}^g+c_{2,2}^g+c_{2,3}^g)(\hat{k}_{1x}\hat{k}_{1y}+\hat{k}_{2x}\hat{k}_{2y}),
\end{split}
\end{equation}
and
\begin{equation}
\begin{split}
&\mathcal{F}_2^{\alpha\beta\gamma,(2)}(\boldsymbol{k}_1,\boldsymbol{k}_2) = 2N_2 P_L(k_1)P_L(k_2)  \\ &\times\left\{(c_{2,1}^g+c_{2,2}^g) \left(\frac{1}{2}\left(\hat{\boldsymbol{k}}_1 \cdot \hat{\boldsymbol{k}}_2\right)\hat{k}_{1y}\hat{k}_{2y} \right)\right. \\  &\hspace{3em}\left.+c_{2,3}^g\left(\frac{1}{4}(\hat{k}_{1y}^2+ \hat{k}_{2y}^2) \right)\right\}.
\end{split}
\end{equation}
The two formalisms for the EFT bispectra can be compared, and from both of them we can write subsets of the whole model, which include the TATT formalism and its extended version with the velocity shear parameter (TATT+VS). The natural components of the 3PCF require the calculation of E and B modes. For $\mapcube$, only E modes suffice due to the construction of the skewness statistics.

\subsection{Tidal alignment and tidal torquing expressions for the IA bispectra}
\label{sec:tatt-ia-bispectrum}
We now write the expressions for the intrinsic alignment bispectra under the TATT formalism. By developing a TATT methodology for the 3PCF in a manner consistent to that usually done for the 2PCF, we restrict ourselves to a well studied smaller parameter space, while verifying the additional constraining power and degeneracy breaking that can come from the addition of third order statistics. Deriving the TATT bispectrum from the EFT parametrization allows us to have a flexible model in which additional terms can be included and investigated separately as needed, such as the case of the velocity-shear parameter, which we describe in Section~\ref{sec:vshear}

From Eqs.~A7 and ~A8 of \citet{bakx2025_2} and from Eq.~67 of \citet{bakx}, we obtain a relation between the EFT parameters $c^g_1$, $c^g_{2,2}$ and $c^g_{2,3}$ and the TATT parameters $C_1$, $C_{1\delta}$, and $C_2$. We assume $c^g_{2,1}=0$, as we will add the velocity shear to our model as a separate contribution.
\begin{equation}
\begin{aligned}
c_1^g &= 2C_1 \\ c_{2,2}^g &= C_2 \\ c_{2,3}^g &= C_{1\delta}-\frac{2}{3}C_2.
\end{aligned}
\end{equation}
We now start with the E mode terms, which will contribute to the mass aperture skewness. We simplify the expressions by writing them, when possible, in terms of the tree-level matter bispectrum
\begin{equation}
\begin{split}
B_{\delta\delta\delta}^{\text{Tree}}(\boldsymbol{k}_1,\boldsymbol{k}_2,\boldsymbol{k}_3) =& 2P_L(k_1)P_L(k_2)F_2(k_1,k_2) \\ &+ 2P_L(k_2)P_L(k_3)F_2(k_2,k_3) \\ &+ 2P_L(k_3)P_L(k_1)F_2(k_3,k_1).
\end{split}
\end{equation}
For the density-density-shape bispectrum, we find the form 
\begin{equation}
\begin{split}
&B_{\delta\delta E}(\boldsymbol{k}_1,\boldsymbol{k}_2,\boldsymbol{k}_3) \\ = &C_1 \left(B^{\text{Tree}}_{\delta\delta\delta}(\boldsymbol{k}_1,\boldsymbol{k}_2,\boldsymbol{k}_3)\right) \\
&+ C_2 P_L(k_1)P_L(k_2)\bigg(\frac{1}{2} (\boldsymbol{\hat{k}}_1 \cdot \boldsymbol{\hat{k}}_2)(\hat{k}_{1x}\hat{k}_{2x}-\hat{k}_{1y}\hat{k}_{2y}) \\ &\hspace{9em}-\frac{1}{12}(\hat{k}_{1x}^2-\hat{k}_{1y}^2+\hat{k}_{2x}^2-\hat{k}_{2y}^2) \bigg) \\
&+ C_{1\delta}P_L(k_1)P_L(k_2) \left(\frac{1}{2}(\hat{k}_{1x}^2-\hat{k}_{1y}^2+\hat{k}_{2x}^2-\hat{k}_{2y}^2)\right).
\end{split}
\label{dde}
\end{equation}
For the density-shape-shape bispectrum, we have 
\begin{equation}
\begin{split}
&B_{\delta EE}(\boldsymbol{k}_1,\boldsymbol{k}_2,\boldsymbol{k}_3) \\
=& C_1^2 \left(B^{\text{Tree}}_{\delta\delta\delta}(\boldsymbol{k}_1,\boldsymbol{k}_2,\boldsymbol{k}_3)\right) \\
&+ \frac{1}{3}C_1C_2U(\boldsymbol{k}_1,\boldsymbol{k}_2,\boldsymbol{k}_3)P_L(k_1)P_L(k_2) \\ 
&+ C_1C_{1\delta}U(\boldsymbol{k}_1,\boldsymbol{k}_2,\boldsymbol{k}_3)P_L(k_1)P_L(k_2)\\
&+ C_1C_2\left(V(\boldsymbol{k}_1,\boldsymbol{k}_2,\boldsymbol{k}_3)-\frac{2}{3}W(\boldsymbol{k}_1,\boldsymbol{k}_2,\boldsymbol{k}_3)\right)P_L(k_1)P_L(k_3)\\ 
&+ C_1C_{1\delta}W(\boldsymbol{k}_1,\boldsymbol{k}_2,\boldsymbol{k}_3)P_L(k_1)P_L(k_3).
\end{split}
\label{bdee}
\end{equation}
where we introduce the auxiliary functions $U$, $V$ and $W$. We thus have
\begin{equation}
\begin{split}
U(\boldsymbol{k}_1,\boldsymbol{k}_2,\boldsymbol{k}_3) &= \frac{1}{2k_3^2}
\Big[k_3^2-(k_1^2+k_2^2)\\&\hspace{4em}\times(1-(\hat{\bm{k}}_1\cdot \hat{\bm{k}}_2)^2)\Big],
\end{split}
\end{equation}
\begin{equation}
\begin{split}
V(\boldsymbol{k}_1,\boldsymbol{k}_2,\boldsymbol{k}_3) &= -\frac{(\hat{\bm{k}}_1 \cdot \hat{\bm{k}}_3)}{2k_3}
\Big[k_2(\hat{\bm{k}}_1 \cdot\hat{\bm{k}}_2) \\ &\hspace{4em} +2k_1(\hat{\bm{k}}_1 \cdot\hat{\bm{k}}_2)^2 -k_1\Big]
\end{split}
\end{equation}
\begin{equation}
\begin{split}
W(\boldsymbol{k}_1,\boldsymbol{k}_2,\boldsymbol{k}_3) = -\frac{1}{2k_3^2}\bigg[ (\boldsymbol{\hat{k}}_1\cdot \boldsymbol{\hat{k}}_2)^2(k_1^2+k_3^2)-k_1^2\bigg].
\end{split}
\end{equation}
For the shape-shape-shape bispectrum, there will be terms proportional to all the permutations of the power spectra products. These terms have two factors of $C_1$ and one factor involving the higher-order parameters. We can write
\begin{equation}
\begin{split}
&B_{EEE}(\boldsymbol{k}_1,\boldsymbol{k}_2,\boldsymbol{k}_3) \\=& C_1^3 \left(B^{\text{Tree}}_{\delta\delta\delta}(\boldsymbol{k}_1,\boldsymbol{k}_2,\boldsymbol{k}_3)\right) \\ 
&+ \frac{1}{6}C_1^2C_2U(\boldsymbol{k}_1,\boldsymbol{k}_2,\boldsymbol{k}_3)P_L(k_1)P_L(k_2) \\
&+ \frac{1}{2}C_1^2C_{1\delta}U(\boldsymbol{k}_1,\boldsymbol{k}_2,\boldsymbol{k}_3)P_L(k_1)P_L(k_2)\\
&+ C_1^2C_2\bigg(T(\boldsymbol{k}_1,\boldsymbol{k}_2,\boldsymbol{k}_3)+\frac{1}{3}W(\boldsymbol{k}_1,\boldsymbol{k}_2,\boldsymbol{k}_3)\bigg)P_L(k_1)P_L(k_3)\\ 
&-\frac{1}{2}C_1^2C_{1\delta}W(\boldsymbol{k}_1,\boldsymbol{k}_2,\boldsymbol{k}_3)P_L(k_1)P_L(k_3)\\
&+ C_1^2C_2\bigg(T(\boldsymbol{k}_2,\boldsymbol{k}_1,\boldsymbol{k}_3)+\frac{1}{3}W(\boldsymbol{k}_1,\boldsymbol{k}_2,\boldsymbol{k}_3)\bigg)P_L(k_2)P_L(k_3)\\ 
&-\frac{1}{2}C_1^2C_{1\delta}W(\boldsymbol{k}_2,\boldsymbol{k}_1,\boldsymbol{k}_3)P_L(k_2)P_L(k_3).
\end{split}
\label{beee}
\end{equation}
where we introduce the auxiliary function $T$. We have
\begin{equation}
\begin{split}
&T(\boldsymbol{k}_1,\boldsymbol{k}_2,\boldsymbol{k}_3) = -\frac{1}{8k_3^2}\bigg[2k_ 3(\boldsymbol{\hat{k}}_1\cdot \boldsymbol{\hat{k}}_3)(k_2(\boldsymbol{\hat{k}}_1\cdot \boldsymbol{\hat{k}}_2)\\
&\hspace{12em}+k_1(2(\boldsymbol{\hat{k}}_1\cdot \boldsymbol{\hat{k}}_2)^2-1))\bigg].
\end{split}
\end{equation}
Following the same methodology, we can write the expressions for the B-mode contributions. The non-vanishing components at tree-level under the TATT approximation are those with a single B-mode. This occurs because only one of the kernels from Eq.~\ref{anisobisp} should go beyond first order, and at first order the B projection gives us $\textbf{M}_{ij}^B \cdot \mathcal{K}^{g,(1)}_{ij}=0$. Therefore, we have
\begin{equation}
\begin{split}
B_{\delta\delta B}(\boldsymbol{k}_1,\boldsymbol{k}_2,\boldsymbol{k}_3) &= \frac{1}{2}(C_2+3C_{1\delta}) \\&\times(\hat{k}_{1x}\hat{k}_{1y}+\hat{k}_{2x}\hat{k}_{2y})P(k_1)P(k_2),
\end{split}
\end{equation}
and
\begin{equation}
\begin{aligned}
B_{\delta EB}(\boldsymbol{k}_1,\boldsymbol{k}_2,\boldsymbol{k}_3) &= \frac{C_1}{4}(C_2+3C_{1\delta}) \\&\times(\hat{k}_{1x}\hat{k}_{1y}+\hat{k}_{2x}\hat{k}_{2y})P(k_1)P(k_2),
\end{aligned}
\label{deb}
\end{equation}

The intrinsic alignment contamination of the cosmic shear signal also includes the permutations between the indices $\delta$, $E$ and $B$ of the bispectra components. In order to compute the permutations, we use Eqs.~\ref{dde}-\ref{deb} with permuted arguments. Since the equations assume $k_3$ to be oriented along the x-axis, each permutation must be accompanied by a rotation of the $\boldsymbol{k}_i$ vectors in order to realign the third argument with the x-axis. For example, we write $B_{\delta E\delta}(\boldsymbol{k}_1,\boldsymbol{k}_2,\boldsymbol{k}_3)$ = $B_{\delta \delta E}(R\boldsymbol{k}_3,R\boldsymbol{k}_1,R\boldsymbol{k}_2)$, where R is the rotation that provides $(R\bm{k}_{2})_y=0$. For the bispectra with one scalar and two shape components, Eqs.~\ref{bdee} and~\ref{deb} assume that the last two arguments refer to the tensor modes. The permutations are followed accordingly.

In order to extend the validity of our modeling to non-linear scales, we follow the phenomenological approach and replace the linear matter power spectrum $P_L(k)$ by the non-linear spectrum $P_{NL}(k)$, as computed from the revised Halofit prescription \cite{Takahashi_2012}. We also note that when taking $C_2=0$ and $C_{1\delta}=0$, our expressions recover the NLA approximation for the IA bispectrum, as used by \citet{Gomes.DES.data}, except for being in terms of the perturbation theory tree-level bispectrum. We thus replace $B^{\text{Tree}}_{\delta\delta\delta}$ on Eqs.~\ref{dde},\ref{bdee} and~\ref{beee} with the non-linear bispectrum computed through the BiHalofit formula \cite{Takahashi.Shirasaki.2019}, which was calibrated from a set of high-resolution cosmological N-body simulations.

\subsection{Extended TATT modeling}
\label{sec:vshear}
To make the TATT model complete at second order, one must add the velocity shear contribution to the perturbative expansion of galaxy shapes. Therefore, Eq.~\ref{tatt_def} becomes:
\begin{equation}
\gamma^I_{ij} = C_1 s_{ij} + C_{1\delta}(\delta s_{ij}) + C_2\left(s_{ik}s_{kj}-\frac{1}{3}\delta_{ij}s^2\right) + C_t t_{ij}.
\label{tatt_vs_def}
\end{equation}
with the $t_{ij}$ tensor relating to the velocity and density fields by $t_{ij} = \hat{S}[\theta-\delta]$, the $\hat{S}[\delta(k)]$ operator being given by
\begin{equation}
\hat{S}[\delta(k)] = \left(\hat{k}_i\hat{k}_j-\frac{1}{3}\delta_{ij}\right)\delta(k)
\end{equation}
and $\theta$ being given by $\theta=\nabla \cdot \boldsymbol{v}$.

Including this contribution to the TATT bispectrum will lead to new additive terms on each of the bispectra components. The contributions to $B_{\delta\delta E}$ and $B_{\delta EE}$ are
\begin{equation}
\begin{split}
    B^{\text{VS}}_{\delta\delta E}(\boldsymbol{k}_1,\boldsymbol{k}_2,\boldsymbol{k}_3) &= - C_tP_L(k_1)P_L(k_2)\\&\times\bigg(\frac{2}{7}(1-(\boldsymbol{\hat{k}}_1\cdot\boldsymbol{\hat{k}}_2))^2 \\&\hspace{2em}+\frac{1}{10}(\hat{k}_{1x}^2-\hat{k}_{1y}^2+\hat{k}_{2x}^2-\hat{k}_{2y}^2) \\&\hspace{2em}+\frac{1}{5}(\boldsymbol{\hat{k}}_1\cdot\boldsymbol{\hat{k}}_2)(\hat{k}_{1x}\hat{k}_{2x}-\hat{k}_{1y}\hat{k}_{2y})\bigg),
\end{split}
\end{equation}
and
\begin{equation}
\begin{split}
    B^{\text{VS}}_{\delta EE}(\boldsymbol{k}_1,\boldsymbol{k}_2,\boldsymbol{k}_3) &=  C_1C_tU_2(\boldsymbol{k}_1,\boldsymbol{k}_2,\boldsymbol{k}_3)P_L(k_1)P_L(k_2) \\ &+ C_1C_tV_2(\boldsymbol{k}_1,\boldsymbol{k}_2,\boldsymbol{k}_3)P_L(k_1)P_L(k_3),
\end{split}
\end{equation}
for which we define the functions $U_2$ and $V_2$ as
\begin{widetext}
\begin{equation}
\begin{split}
U_2(\boldsymbol{k}_1,\boldsymbol{k}_2,\boldsymbol{k}_3) &= -\frac{1}{70k_1k_2k_3^4}
\bigg[7(\boldsymbol{\hat{k}}_1 \cdot \boldsymbol{\hat{k}}_2)(k_1^6+k_2^6-k_3^2(k_1^4+k_ 2^2)+9(k_1^2k_2^4+k_1^4k_2^2)+2k_1^2k_2^2k_3^2) \\ &\hspace{8em}+ (10+32(\boldsymbol{\hat{k}}_1 \cdot \boldsymbol{\hat{k}}_2)^2)(k_1^5k_2+k_2^5k_1+2k_1^3k_2^3) \\&\hspace{8em}+2(\boldsymbol{\hat{k}}_1 \cdot \boldsymbol{\hat{k}}_2)^2(20k_1^3k_2^3-14k_1k_2k_3^2(k_1^2+k_2^2))\\ &\hspace{8em}+ 2k_1^2k_2^2(\boldsymbol{\hat{k}}_1 \cdot \boldsymbol{\hat{k}}_2)^3(22k_1^2+22k_2^2-k_3^2)+16(\boldsymbol{\hat{k}}_1 \cdot \boldsymbol{\hat{k}}_2)^4k_1^3k_2^3\bigg],
\end{split}
\end{equation}
\begin{equation}
\begin{aligned}
V_2(\boldsymbol{k}_1,\boldsymbol{k}_2,\boldsymbol{k}_3) = \frac{2}{5}\frac{k_1}{k_3}(\boldsymbol{\hat{k}}_1 \cdot \boldsymbol{\hat{k}}_3)((\boldsymbol{\hat{k}}_1 \cdot \boldsymbol{\hat{k}}_2)^2-1)-\frac{1}{35}(5+2(\boldsymbol{\hat{k}}_1 \cdot \boldsymbol{\hat{k}}_3)^2).
\end{aligned}
\end{equation}
\end{widetext}
The velocity-shear contribution to the shape-shape-shape component can be written in terms of the same functions as
\begin{equation}
\begin{aligned}
        B^{\text{VS}}_{EEE}(\boldsymbol{k}_1,\boldsymbol{k}_2,\boldsymbol{k}_3) &=  \frac{1}{2}C_1^2C_tU_2(\boldsymbol{k}_1,\boldsymbol{k}_2,\boldsymbol{k}_3)P_L(k_1)P_L(k_2) \\ &+ \frac{1}{2}C_1^2C_tV_2(\boldsymbol{k}_1,\boldsymbol{k}_2,\boldsymbol{k}_3)P_L(k_1)P_L(k_3) \\ &+ \frac{1}{2}C_1^2C_tV_2(\boldsymbol{k}_2,\boldsymbol{k}_1,\boldsymbol{k}_3)P_L(k_2)P_L(k_3).
\end{aligned}
\end{equation}
Finally, the additional contributions to the B-mode bispectra are of a similar form as the standard TATT contributions. We write:

\begin{equation}
\begin{split}
B_{\delta\delta B}^{\text{VS}}(\boldsymbol{k}_1,\boldsymbol{k}_2,\boldsymbol{k}_3) &= -\frac{4}{5}C_t(\hat{k}_{1x}\hat{k}_{1y}+\hat{k}_{2x}\hat{k}_{2y}) \\ &\times P(k_1)P(k_2),
\end{split}
\end{equation}
and
\begin{equation}
\begin{split}
B_{\delta EB}^{\text{VS}}(\boldsymbol{k}_1,\boldsymbol{k}_2,\boldsymbol{k}_3) =  &-\frac{2}{5}C_1C_t(\hat{k}_{1x}\hat{k}_{1y}+\hat{k}_{2x}\hat{k}_{2y}) \\ &\times P(k_1)P(k_2).
\end{split}
\end{equation}

\subsection{The convergence bispectrum}

Finally, we can compute the TATT contamination for the convergence bispectra, which includes the combinations $B_{\kappa\kappa E}$, $B_{\kappa EE}$, $B_{EEE}$, and their permutations of $\kappa$ and E indices. The lensing window function is given in terms of the lensing efficiency by
\begin{equation}
W_{\kappa}(\chi) = \frac{3\Omega_m H_0^2}{2c} \frac{\chi}{a(\chi)}q(\chi).
\end{equation}

Therefore, under the Limber approximation, we use the components from Eqs.~\ref{dde},~\ref{bdee} and~\ref{beee} and write
\begin{equation}
\begin{aligned}
B^{\kappa \kappa E}_{ijk}(\boldsymbol{l}_1, \boldsymbol{l}_2, \boldsymbol{l}_3) 
 &= \int d\chi \frac{W_{\kappa}^i(\chi)W_{\kappa}^j(\chi) W_g^k(\chi) }{\chi^4} \\ &\hspace{4em}\times B_{\delta\delta E}(\boldsymbol{l}_1/\chi, \boldsymbol{l}_2/\chi, \boldsymbol{l}_3/\chi)
 \label{eqs.kke}
\end{aligned}
\end{equation}
\begin{equation}
\begin{split}
B^{\kappa EE}_{ijk}(\boldsymbol{l}_1, \boldsymbol{l}_2, \boldsymbol{l}_3) 
&= \int d\chi \frac{W_{\kappa}^i(\chi) W_g^j(\chi)W_g^k(\chi) }{\chi^4}  \\ &\hspace{4em}\times B_{\delta EE}(\boldsymbol{l}_1/\chi, \boldsymbol{l}_2/\chi, \boldsymbol{l}_3/\chi)
\label{eqs.kee}
\end{split}
\end{equation}
\begin{equation}
\begin{split}
B^{EEE}_{\text{proj}}(\boldsymbol{l}_1, \boldsymbol{l}_2, \boldsymbol{l}_3) 
&= \int d\chi \frac{W_g(\chi)^iW_g(\chi)^jW_g(\chi)^k}{\chi^4} \times \\ &\hspace{4em}\times B_{EEE}(\boldsymbol{l}_1/\chi, \boldsymbol{l}_2/\chi, \boldsymbol{l}_3/\chi)
\label{eqs.eee}
\end{split}
\end{equation}

For the permutations $B^{\kappa E \kappa}_{ijk}$, $B^{E \kappa\kappa}_{ijk}$, $B^{E\kappa E}_{ijk}$, and $B^{EE \kappa}_{ijk}$, the bin indices must be permuted accordingly on the window functions, and the bispectra must be swapped by their permuted counterparts.

\subsection{The three-point correlation function}
We now compute the TATT contamination signal on the shear three-point correlation function. Since we will be compressing the full 3PCF information into the mass aperture statistic, we only require the use of E modes. A computation of the B mode contributions to the full 3PCF can be useful to understand potential nonzero signal on data. We leave this for future study. We consider the three-point correlation function of three distinct fields A, B, C at $\bm{X}_1$,$\bm{X}_2$, and $\bm{X}_3$. We write the natural components in the $\times$ projection, as defined by \citet{sugiyama2024fastmodelingshearthreepoint}. Next, we write the component $\Gamma_0$ as
\begin{widetext}
\begin{align}
    \Gamma_0^{ABC}(\theta_1, \theta_2, \varphi)
    &= - \langle
    \gamma_{\rm c}^A(\bm{X}_1)
    \gamma_{\rm c}^B(\bm{X}_2)
    \gamma_{\rm c}^C(\bm{X}_3)
    \rangle 
    e^{-3i(\varphi_1+\varphi_2)} \\
    &= -\int_{\bm{\ell}_1, \bm{\ell}_2,\bm{\ell}_3}
    \langle
    \gamma_{\rm c}^A(\bm{\ell}_3)
    \gamma_{\rm c}^B(\bm{\ell}_1)
    \gamma_{\rm c}^C(\bm{\ell}_2)
    \rangle
    e^{-i(\bm{\ell}_3\cdot\bm{X}_1+\bm{\ell}_1\cdot\bm{X}_2+\bm{\ell}_2\cdot\bm{X}_3)}
    e^{-3i(\varphi_1+\varphi_2)} \\
    &= -\int_{\bm{\ell}_1, \bm{\ell}_2,\bm{\ell}_3}
    \langle
    \kappa_{\rm c}^A(\bm{\ell}_3)
    \kappa_{\rm c}^B(\bm{\ell}_1)
    \kappa_{\rm c}^C(\bm{\ell}_2)
    \rangle
    e^{-2i\sum\beta_i}
    e^{-i(\bm{\ell}_3\cdot\bm{X}_1+\bm{\ell}_1\cdot\bm{X}_2+\bm{\ell}_2\cdot\bm{X}_3)}
    e^{-3i(\varphi_1+\varphi_2)} \\
    &= -\int_{\bm{\ell}_1, \bm{\ell}_2,\bm{\ell}_3}
    B_{ABC}(\bm{\ell}_3, \bm{\ell}_1, \bm{\ell}_2)\delta\left(\sum\bm{\ell}_i\right)
    e^{-2i\sum\beta_i}
    e^{-i(\bm{\ell}_3\cdot\bm{X}_1+\bm{\ell}_1\cdot\bm{X}_2+\bm{\ell}_2\cdot\bm{X}_3)}
    e^{-3i(\varphi_1+\varphi_2)} \\
    &= -\int_{\bm{\ell}_1, \bm{\ell}_2,\bm{\ell}_3}
    B_{ABC}(\bm{\ell}_3, \bm{\ell}_1, \bm{\ell}_2)
    e^{-2i\sum\beta_i}
    e^{-i(\bm{\ell}_1\cdot\bm{\theta}_1 + \bm{\ell}_2\cdot\bm{\theta}_2)}
    e^{-3i(\varphi_1+\varphi_2)},
\end{align}
\end{widetext}
where we perform the multipole expansion of the bispectrum opening angle by
\begin{equation}
\begin{aligned}
B_{ABC}(\bm{\ell}_3, \bm{\ell}_1, \bm{\ell}_2) &=b_{BCA}(\ell_1,\ell_2, \alpha) \\&= \sum_{L=0}^\infty b^L_{BCA}(\ell_1,\ell_2)P_L(\cos\alpha).
\end{aligned}
\end{equation}
Here we use the Legendre expansion because the E-mode TATT bispectra are parity-even. The choice of expansion is dependent on the parity of the bispectrum. For the B-mode contribution, which is parity-odd, we must perform a separate expansion into parity-odd functions. Since the $\mapcube$ computation only includes contributions from E-modes, we restrict ourselves to the E-modes of the 3PCF, and thus opt for the Legendre expansion.

The multipoles are given by
\begin{equation}
\begin{aligned}
b^L_{ABC}(\ell_1,\ell_2) &= \frac{2L+1}{2} \\ &\times \int_{-1}^1\text{d}(\cos\alpha)b_{ABC}(\ell_1,\ell_2,\alpha)P_L(\cos\alpha).
\end{aligned}
\end{equation}
The three-point correlation function for the fields A, B and C can be expanded by
\begin{equation}
\Gamma^{ABC}_0(\theta_1, \theta_2,\phi) = \frac{1}{2\pi}\sum_{M=-\infty}^\infty e^{iM\phi}\Gamma^{ABC, M}_0(\theta_1,\theta_2)
\end{equation}
Now we can write each of the 3PCF multipoles as
\begin{equation}
\begin{aligned}
\Gamma^{ABC, M}_0(\theta_1,\theta_2)&=\frac{1}{(2\pi)^3}\int \text{d}\ln\ell_1\text{d}\ln\ell_2 \\&\hspace{5em}\times \ell_1^2\ell_2^2H^{ABC}_M(\ell_1,\ell_2) \\ &\hspace{5em}\times J_{M-3}(\ell_1\theta_1)J_{-M-3}(\ell_2\theta_2),
\end{aligned}
\end{equation}
where the kernel function $H^{ABC}_M(\ell_1,\ell_2)$ is given in terms of the bispectrum multipoles and of the cosmology independent multipole coupling functions $G_{LM}(\psi)$. We have
\begin{equation}
H^{ABC}_M(\ell_1,\ell_2) = \sum_{L=0}^{\infty}(-1)^L b^{ABC}_L(\ell_1,\ell_2)
\end{equation}
We now consider the case where the A, B and C indices can each assume the value of either $\kappa$ or E. We compute all the possible permutations, following Eqs.~\ref{eqs.kke},~\ref{eqs.kee}, and~\ref{eqs.eee}, and sum them to arrive at the total contamination signal to the bispectrum for each tomographic bin combination. Next, following the procedure detailed in \citet{sugiyama2024fastmodelingshearthreepoint}, we compute the multipoles of the bispectrum on an FFT grid in $\ell_1$ and $\ell_2$. We also analogously compute the remaining components $\Gamma^1$, $\Gamma^2$, and $\Gamma^3$. To increase the computational efficiency, we mediate this calculation with a previous computation and interpolation of the total intrinsic alignment bispectrum on a 3D grid. In this case, the ordering of the triangle sides is important because each side refers to a different tracer. Thus, we introduce a parametrization that is sensitive to the permutation of triangle sides.

We define the variables $p$, $u$ and $v$ by
\begin{equation}
\begin{cases}
p&=\ell_1+\ell_2+\ell_3, \\ u &= \ell_1/p, \\ v &= \ell_2/p.
\end{cases}
\end{equation}
We vary the values of $p$ across several orders of magnitude to encompass triangles of different scales, while $u$ and $v$ are used to compute the triangle shape. 

Finally, after arriving at the bispectra multipole components, we use a double Hankel transformation to arrive at the 3PCF multipoles, which are then summed to give the complete three-point correlation function. Thus, we arrive at the total E-mode contamination to the natural components of the three-point correlation function. In this process, we truncate the multipole expressions at $L_{\max}=50$ and $M_{\max}=100$.

\subsection{The mass aperture skewness}

\begin{figure*}
    \centering
    \includegraphics[width=0.95\linewidth]{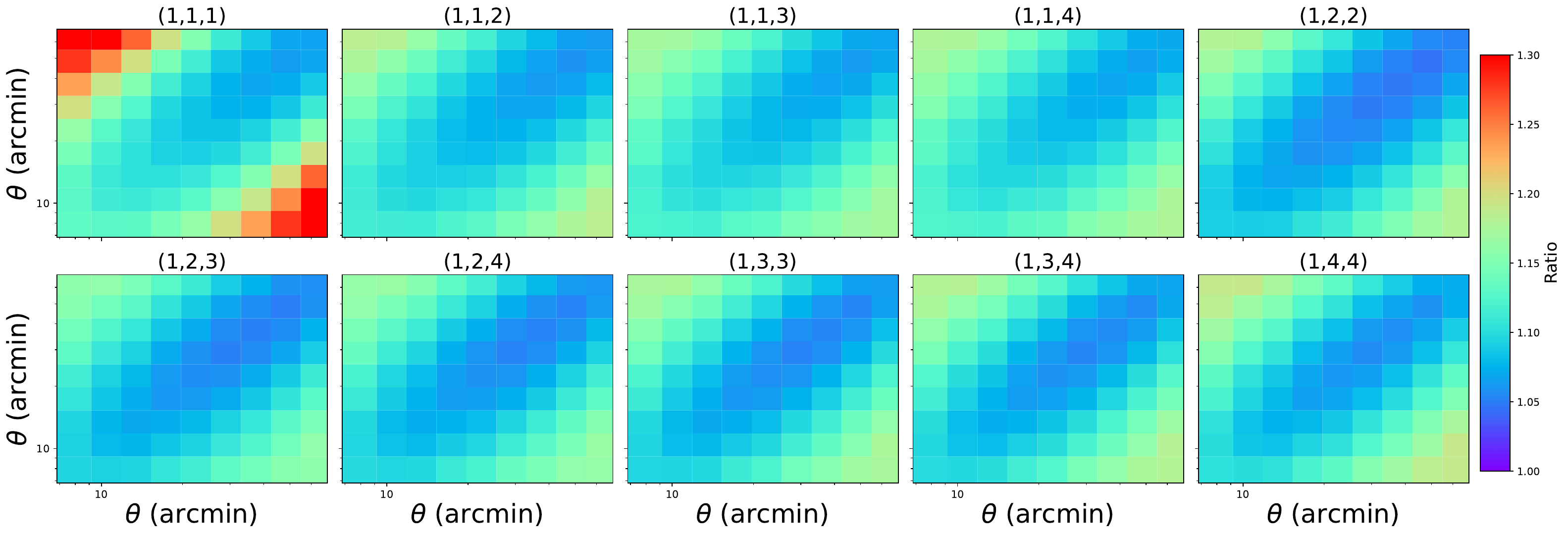}
    \includegraphics[width=0.95\linewidth]{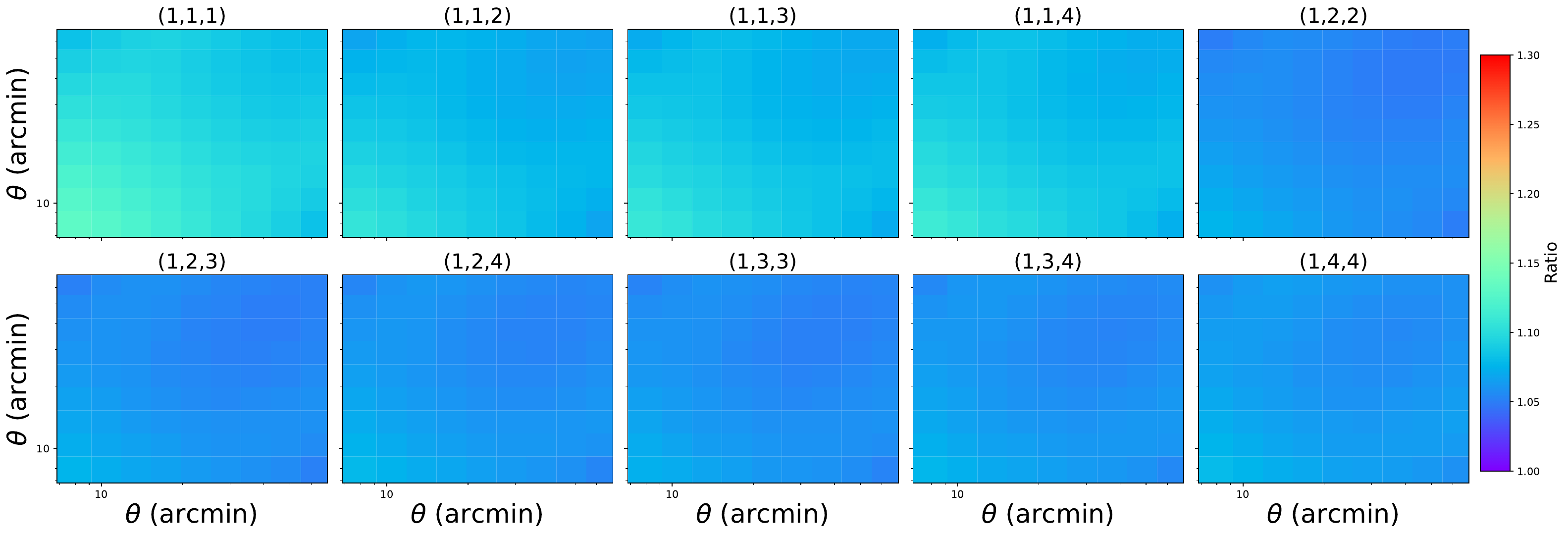}
    \caption{Ratio between the theoretical shear 3PCF signal with intrinsic alignment contamination and the pure cosmic shear signal. Here we show the real part of $\Gamma^0$. The upper panels use our TATT modeling, while the lower panels use NLA modeling. Here we only show bin combinations that include at least one instance of the lowest redshift bin. The alignment and torquing amplitudes are taken from the best fit of the DES Y3 shear analysis. The x and y axes show two of the triangle sides, where we fix the angle between them at $\phi=60\deg$. The color bars in the upper and lower panels are the same. The TATT contamination varies more strongly across different triangle configurations than the NLA prediction.}
    \label{fig:3pcfTATT}
\end{figure*}

\begin{figure*}
    \centering
    \includegraphics[width=0.95\linewidth]{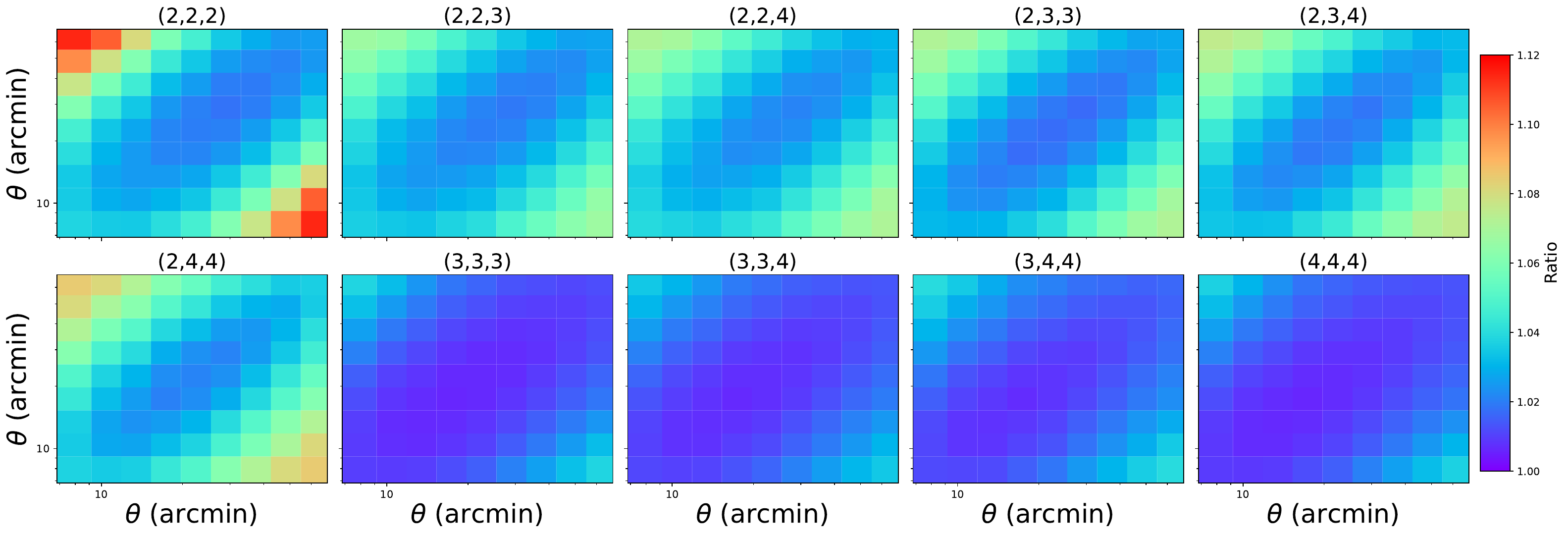}
    \includegraphics[width=0.95\linewidth]{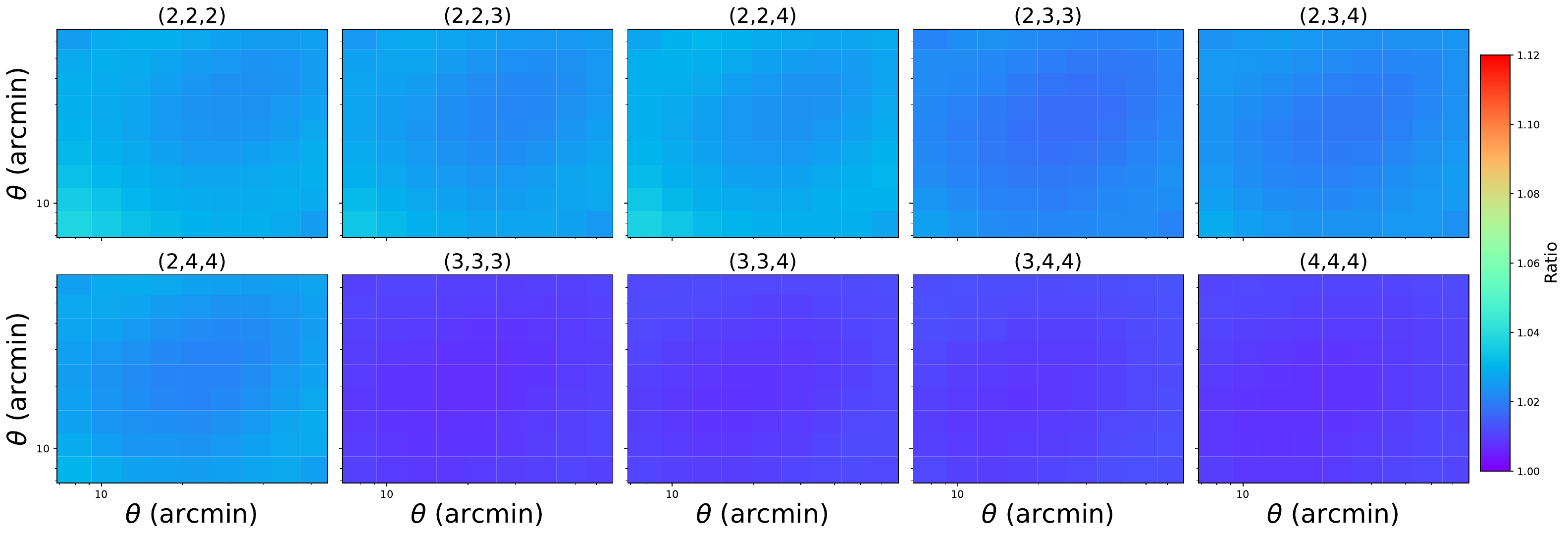}
    \caption{
    Same plot as Fig.~\ref{fig:3pcfTATT}, but here we show the 3PCF only for the bin combinations that do not include any instance of the lowest redshift bin, thus having a significantly lower IA contamination. We show the real part of $\Gamma^0$. The upper panels use our TATT modeling, while the lower panels use NLA modeling.  The alignment and torquing amplitudes are taken from the best fit of the DES Y3 shear analysis. The x and y axes show two of the triangle sides, where we fix the angle between them at $\phi=60\deg$. The color bars in the upper and lower panels are the same.}
    \label{fig:3pcfTATT2}
\end{figure*}

Now we consider the computation of the mass aperture statistic from the natural components of the 3PCF. With $\Gamma^{\text{TATT}}_i$ being the total E-mode contamination to each natural component, we write
\begin{equation}
\begin{split}
&\langle\mathcal{M}_{\rm ap}^3\rangle^{\text{TATT}}(\theta) = 
\frac{3}{2}\text{Re}\left[\int \frac{sds}{s^2}\int_{s<t'<|t'-s|} \frac{d^2\bm{t}'}{2\pi\theta^2} \right.\\ 
&\hspace{3em}\times \left.\sum_{i=0,1,2,3}\Gamma^{\text{TATT}}_i(s,\bm{t}')T_i\left(\frac{s}{\theta},\frac{\bm{t}'}{\theta}\right)
\right],
\end{split}
\end{equation}
where the $T_0$ and $T_1$ functions are given by Eqs.~51-52 of \citet{Jarvis.Jain.2003}, and the quantities s and $\textbf{t'}$ are given in their Eqs.~46-48 in terms of the vectors $\textbf{q}_i$ connecting each vertex of a triangle configuration to the triangle centroid. The functions $T_2$ and $T_3$ are found by permuting the $\textbf{q}_i$ indices in the equation for $T_1$. This gives us a total $\mapcube^{\text{TATT}}$ term that must be added to the pure shear $\mapcube$ to model the observed signal.

In order to incorporate the non-linear extension described in Section~\ref{sec:tatt-ia-bispectrum}, we separate the seven additive IA contributions to the total shear $\mapcube$, by writing each of $\mapcube_{ABC}(\theta)$ separately, with A, B, and C each being either $\delta$ or E. We compute each of these contributions without their terms proportional to $B^{\text{Tree}}_{\delta\delta\delta}$. Then, we compute the pure shear $\mapcube$ from the BiHalofit matter bispectrum, adding the tidal alignment contribution directly into the lensing kernel as done in Eq.~\ref{nlakernel}. This procedure leaves us with eight terms that contribute to the observed $\mapcube$ signal. The first includes shear and first order IA effects. The other seven terms include the higher-order IA effects.
We also allow for the computation of the redshift-dependent mass aperture $\mapcube(\theta, z)$, which is ideal for cosmological inference because it is a quantity independent of redshift distribution functions and shear calibration nuisance parameters. Therefore, we perform the bispectrum multipole expansion, FFT, and conversion to $\mapcube$ directly from the matter bispectrum and from the quantities defined by Eqs.~\ref{dde},~\ref{bdee} and~\ref{beee}, prior to line-of-sight integration. Then, we perform the Limber approximation integrals as the last step of the calculation. With this method, it is possible to train emulator models for the $\mapcube(\theta, z)$ components and use them in diverse inference scenarios. In summary, we obtain
\begin{equation}
\begin{split}
&\mapcube^{\text{Total}}_{ijk}(\theta) = \int d\chi \\ &\times \Bigg[\frac{W^i_{\kappa}(\chi)W^j_{\kappa}(\chi)W^k_{\kappa}(\chi)}{\chi^4} \mapcube_{\delta\delta\delta}(\theta, z(\chi)) \\ &\hspace{2em}+ \frac{W^i_{\kappa}(\chi)W^j_{\kappa}(\chi)W^k_{g}(\chi)}{\chi^4} \mapcube_{\delta\delta E}(\theta, z(\chi))  \\ &\hspace{2em}+\frac{W^i_{\kappa}(\chi)W^j_{g}(\chi)W^k_{\kappa}(\chi)}{\chi^4} \mapcube_{\delta E \delta}(\theta, z(\chi))  \\ &\hspace{2em}+\frac{W^i_{g}(\chi)W^j_{\kappa}(\chi)W^k_{\kappa}(\chi)}{\chi^4} \mapcube_{E \delta \delta}(\theta, z(\chi)) \\ &\hspace{2em}+\frac{W^i_{\kappa}(\chi)W^j_{g}(\chi)W^k_{g}(\chi)}{\chi^4} \mapcube_{\delta EE}(\theta, z(\chi)) \\ &\hspace{2em}+\frac{W^i_{g}(\chi)W^j_{\kappa}(\chi)W^k_{g}(\chi)}{\chi^4} \mapcube_{E\delta E}(\theta, z(\chi)) \\ &\hspace{2em}+\frac{W^i_{g}(\chi)W^j_{g}(\chi)W^k_{\kappa}(\chi)}{\chi^4} \mapcube_{E E \delta}(\theta, z(\chi)) \\ &\hspace{2em}+\frac{W^i_{g}(\chi)W^j_{g}(\chi)W^k_{g}(\chi)}{\chi^4} \mapcube_{EEE}(\theta, z(\chi))\Bigg],
\end{split}
\end{equation}
where the $W_{\kappa}$ functions are computed from the NLA-modified lensing kernels.

\section{Results}

\subsection{Intrinsic alignment contamination to the three-point correlation function}

We use the best fit cosmological and intrinsic alignment parameters from the Dark Energy Survey Year 3 cosmic shear analysis by \citet{Secco.Samuroff} to generate realistic theoretical predictions of the shear three-point correlation function under the NLA and TATT models. We compute $\Gamma_0$, $\Gamma_1$, $\Gamma_2$, and $\Gamma_3$. In Figures~\ref{fig:3pcfTATT} and \ref{fig:3pcfTATT2}, we show the ratio of $\Gamma_0$ computed with and without intrinsic alignment contamination. The upper panels use the TATT model, while the lower panels use NLA. The parameter values are listed in Table~\ref{table:seccoparams}. While the DES Y3 values indicate low IA amplitudes, other studies have found larger values suggesting a higher contamination level. Our choice for fiducial values is, therefore, conservative.

Figure~\ref{fig:3pcfTATT} shows the results for redshift bin combinations that include the lowest z bin. We find that the range of values for the contamination level is higher for the TATT model, achieving $20\%-30\%$ for some of the triangle configurations. The structure of the off-isosceles cases is most clearly seen with the TATT model, showing that the higher-order terms upweight the more skewed configurations. In contrast, the NLA prediction captures only the contribution that is more homogeneous across triangle configurations. The contribution of the IA effect to isosceles triangles, therefore, is similar between NLA and TATT, being at the level of $10\%$ for both models.

For redshift bin combinations that exclude the lowest z bin, the effect of IA is more modest for both NLA and TATT. While we still see that TATT introduces an additional structure to the triangle configuration dependence of the 3PCF, the contamination level is almost always below $10\%$ for both TATT and NLA. 
\begin{table}[h]
    \centering
    \caption{Cosmological and intrinsic alignment parameters used for predictions of the three-point correlation function. Values are taken from the mean posteriors of the DES Y3 $\Lambda$CDM cosmic shear analysis by \citet{Secco.Samuroff}}.
    \label{table:seccoparams}
    \begin{tabular}{||c|c||}
    \hline
    Parameter & value \\
    \hline
    \hline
    $\Omega_{\rm m}$ & 0.289 \\ 
    $S_8$ & 0.772  \\
    $h_0$ & 0.722  \\
    $n_s$ & 0.959 \\
    $\Omega_b$ & 0.0463  \\
    \hline
    $A_1$ & -0.33 \\
    $\alpha_1$ & 2.81 \\
    $A_2$ & 0.65 \\
    $\alpha_2$ & 1.69 \\
    $b_{\text{TA}}$ & 0.91 \\
    \hline
    \end{tabular}
    \vspace{0.5cm}
\end{table}

\subsection{Intrinsic alignment contamination to the mass aperture skewness}

\begin{figure*}
    \centering  
    \includegraphics[width=0.48\textwidth]{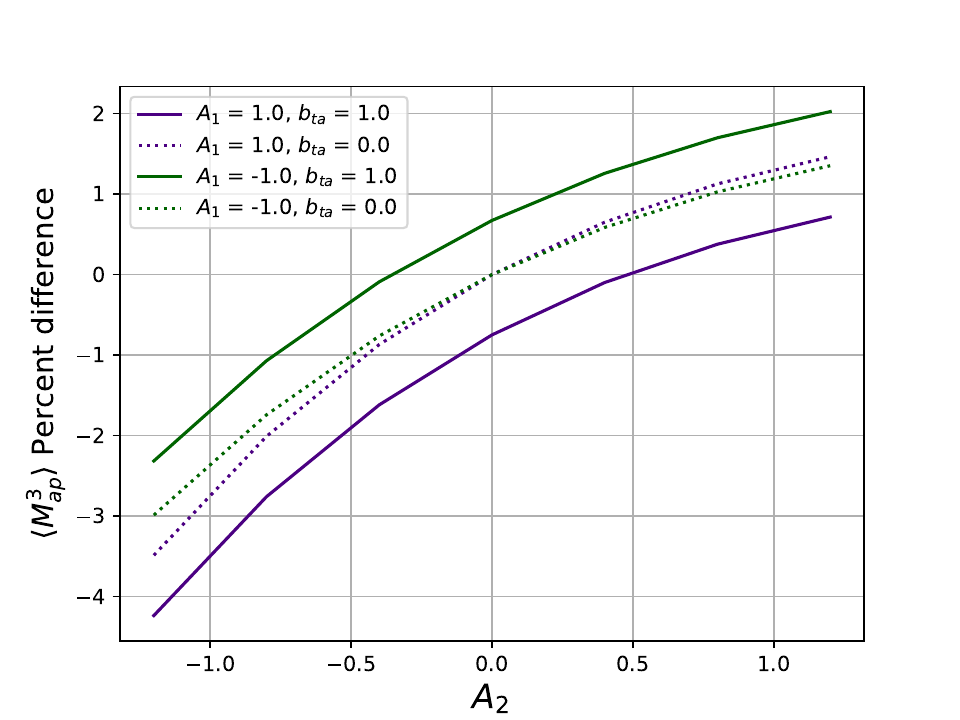}
    \includegraphics[width=0.48\textwidth]{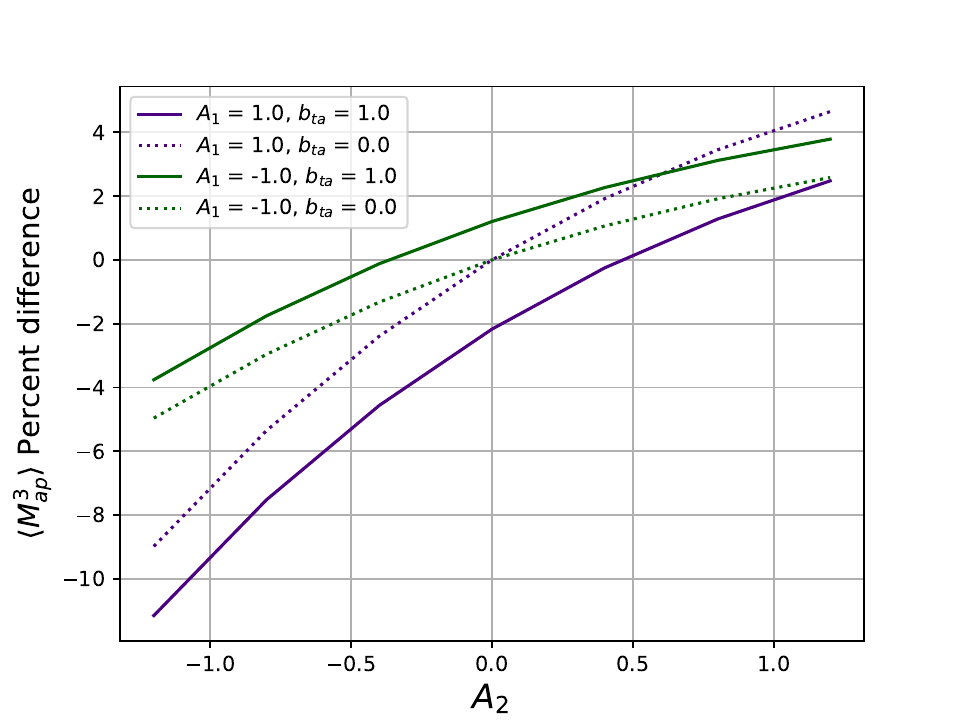}
    \caption{Percent difference between the total $\mapcube$ signal computed with a TATT and with an NLA intrinsic alignment contamination. We show how the tidal torquing effect described by the $A_2$ parameter and the inclusion of density weighting shift the theoretical prediction of the mass aperture skewness. The left left panel shows the results for bin combination $(1,1,1)$, while the right panel shows the results for bin combination $(1,3,4)$. We fix our filter aperture radius at $\theta=14'$. For different filters, the parameter dependence is similar, but the overall IA contamination is smaller. We see that the difference between NLA and TATT can approch $10\%$ of the total $\mapcube$ signal for $A_1=1$, $A_2=-1$, $b_{\text{TA}}=1.0$ in the case of bin combination $(1,3,4)$.
    }
    \label{varya2}
\end{figure*}

\begin{figure*}
    \centering  
    \includegraphics[width=0.48\textwidth]{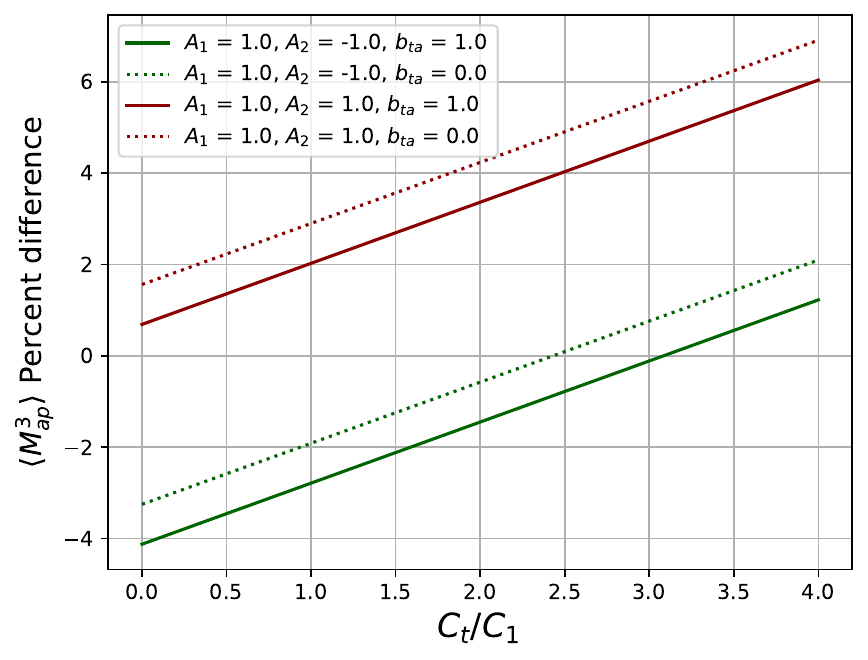}
    \includegraphics[width=0.48\textwidth]{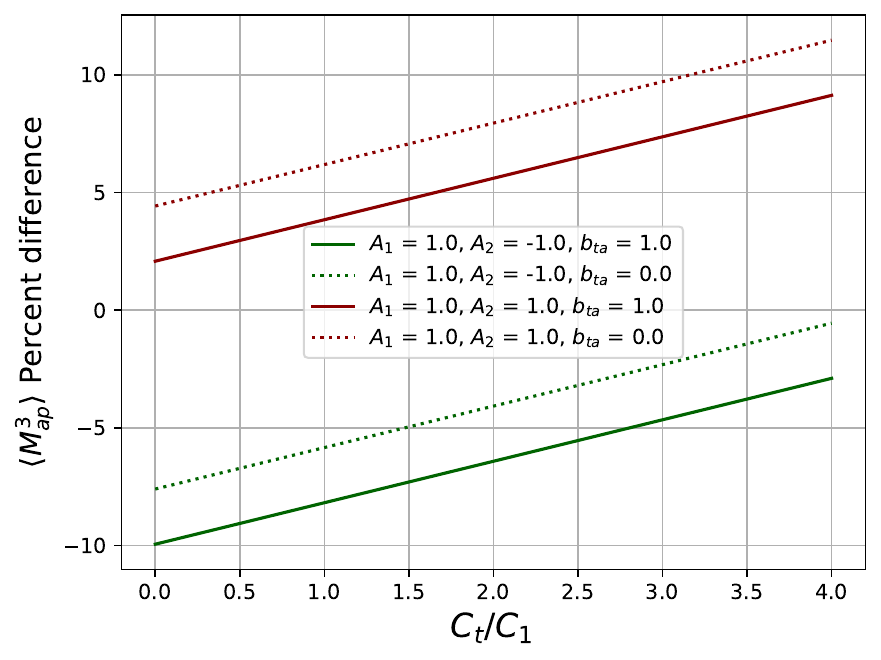}
    \caption{Percent difference between the total $\mapcube$ signal computed with a TATT+VS and with an NLA intrinsic alignment contamination. We show how the velocity shear strength modulated by the $C_t$ parameter shifts the theoretical prediction of the mass aperture skewness. The left left panel shows the results for bin combination $(1,1,1)$, while the right panel shows the results for bin combination $(1,3,4)$. We fix our filter aperture radius at $\theta=14'$. The impact of the a velocity shear term with magnitude $C_t=4C_1$ is that of a shift of around $5\%-8\%$ on the total $\mapcube$.  
    }
    \label{varyct}
\end{figure*}

\begin{figure*}
    \centering
    \includegraphics[width=0.48\textwidth]{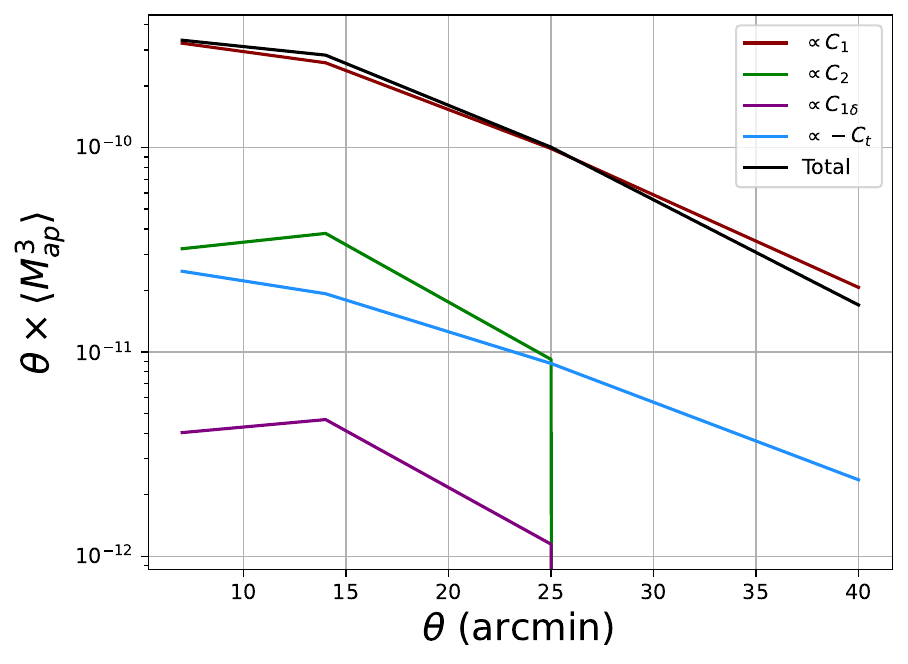}
    \includegraphics[width=0.48\textwidth]{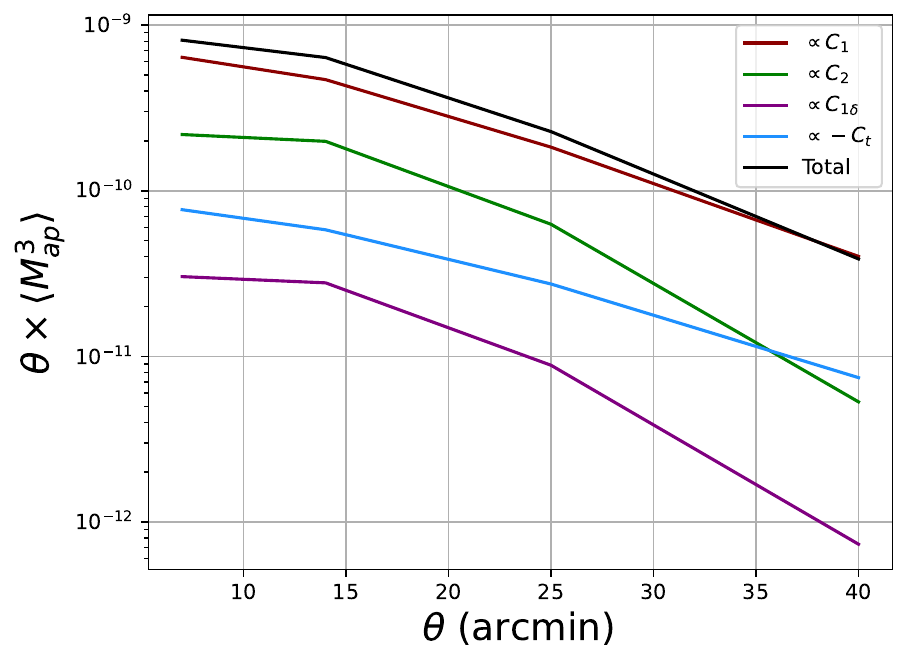}
    \caption{Individual contributions of the perturbative expansion terms to the total $\mapcube$. The red line is the sum of the first order terms, which are proportional to $C_1$, $C_1^2$ or $C_1^3$, depending on how many shape fields are included. The green line sums the terms proportional to $C_2$, $C_1C_2$, and $C_1^2C_2$. The purple line includes terms that are proportional to $C_{1\delta}$, $C_1C_{1\delta}$, and $C_1^2C_{1\delta}$. Finally, the light blue line includes terms proportional to $C_t$ and its combinations with powers of $C_1$, which are multiplied by -1 for ease of visualization. The black line indicates the total intrinsic alignment $\mapcube$ signal. The left panel shows the results for bin combination $(1,1,1)$, while the right panel shows the results for bin combination $(1,3,4)$.}
    \label{fig:paramcontributions}
\end{figure*}

\begin{figure*}
    \centering
    \includegraphics[width=0.48\textwidth]{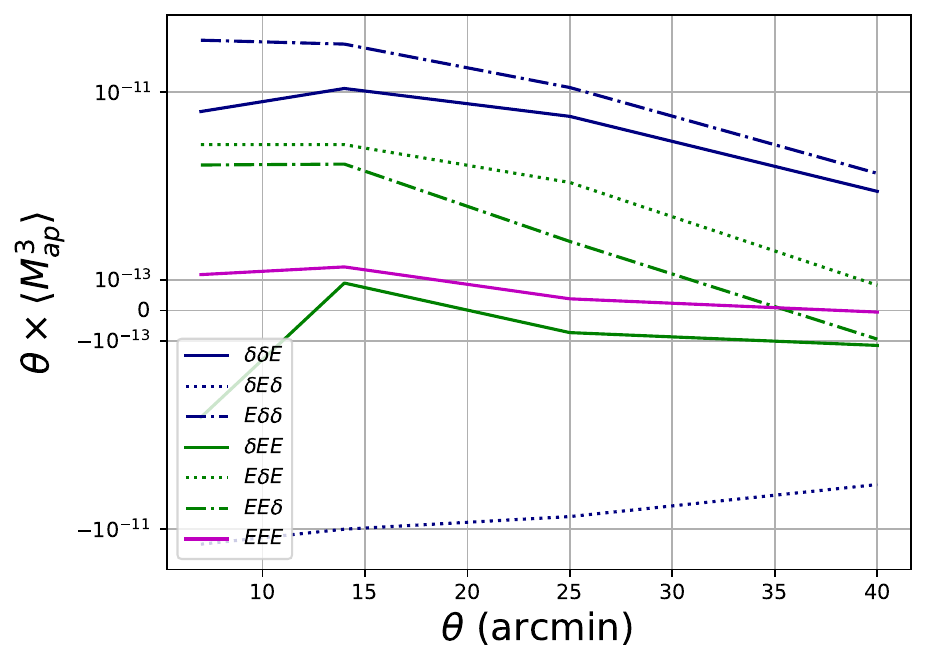}
    \includegraphics[width=0.48\textwidth]{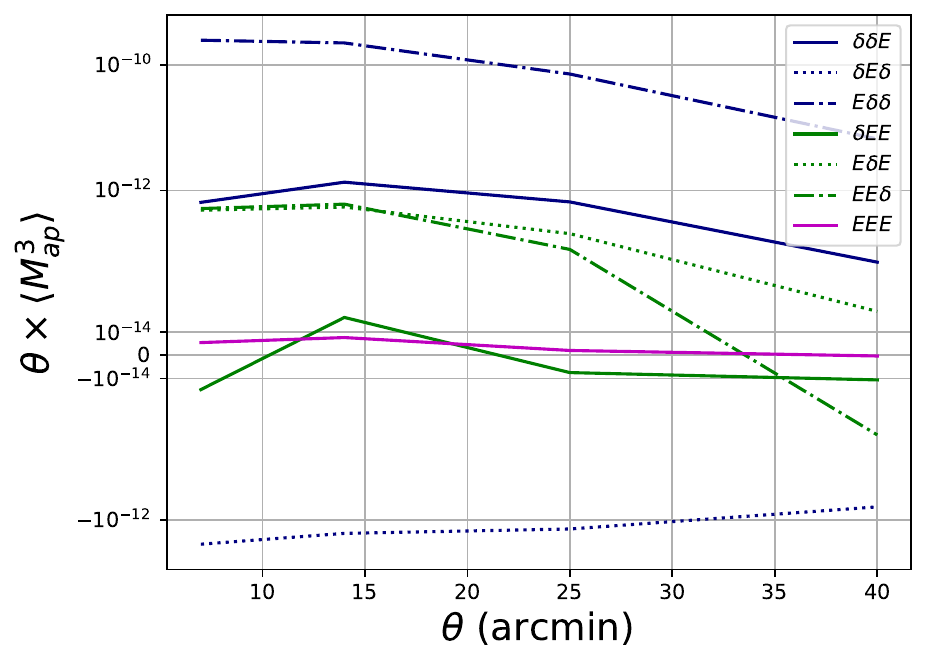}
    \caption{Contributions of the bispectra with different combinations of fields to the total $\mapcube$. The left panel shows the results for bin combination $(1,1,1)$, while the right panel shows the results for bin combination $(1,3,4)$. In blue we show combinations with two density fields and one shape field. The combinations in green have two shape fields and one density field. The magenta line is the term with only shape fields.}
    \label{fig:bispeccontributions}
\end{figure*}

\begin{figure*}
    \centering
    \includegraphics[width=\linewidth]{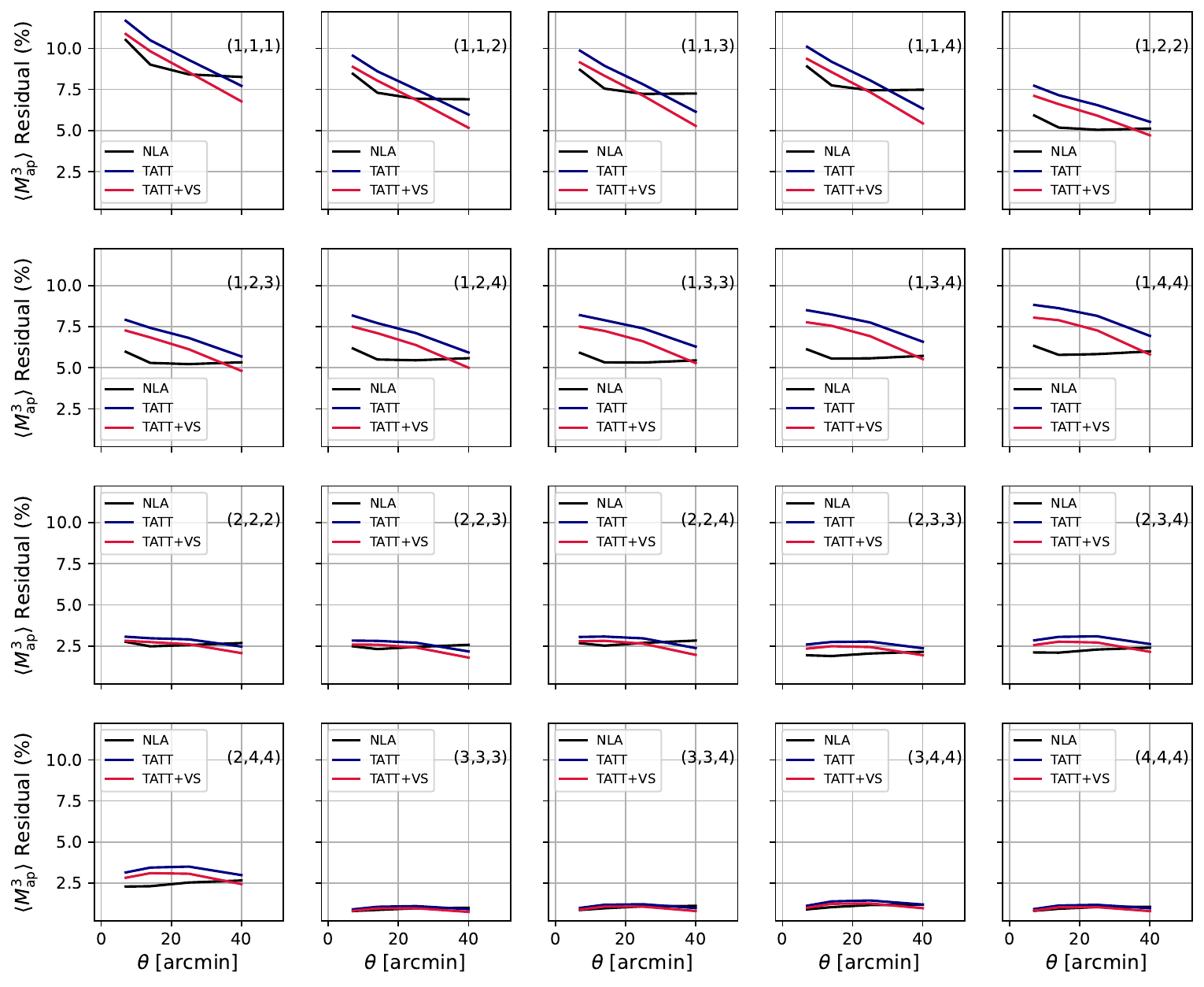}
    \caption{Percent value of the total mass aperture skweness residual relative to a baseline model with no intrinsic alignment. The blue (black) line shows the computation based on the TATT (NLA) model. For the red line, we include the velocity shear term on the TATT computation, assuming its value is given by the Lagrangian co-evolution relation. The difference between TATT+VS and TATT is of the order of $1\%$.}
    \label{fig:map3IA}
\end{figure*}

\begin{figure*}
    \centering
    \includegraphics[width=\linewidth]{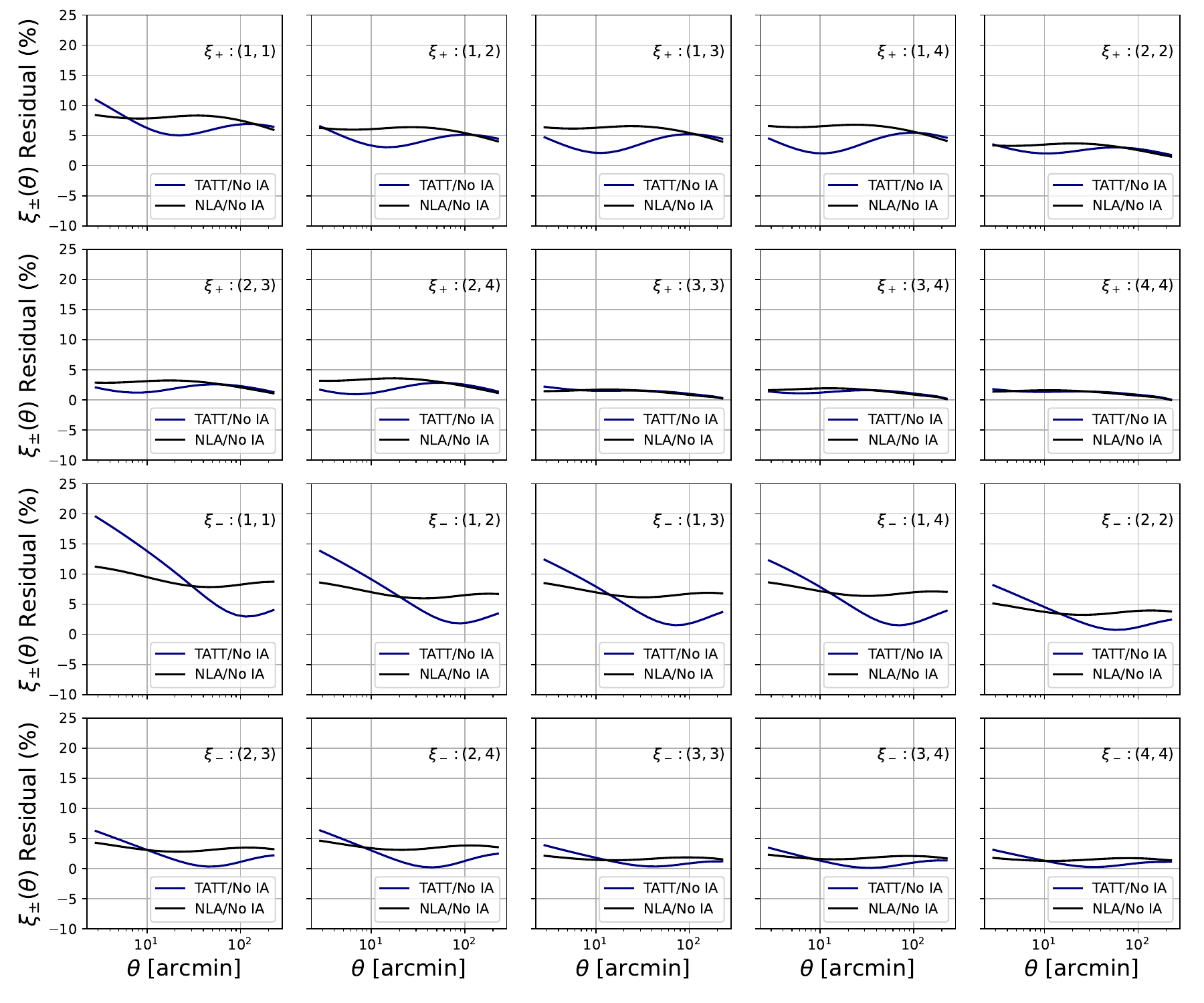}
    \caption{The blue (black) line shows the residual of the cosmic shear signal predicted with the IA contamination based on TATT (NLA) relative to the signal without any IA contamination. The alignment and torquing amplitudes are taken from the best fit of the DES Y3 shear analysis. The impact of changing between NLA and TATT is larger for the three-point correlation function than for the two-point function. For the mass aperture skewness, the impact is more modest in magnitude, while carrying an opposite sign for most scales of interest.}
    \label{fig:2pcfIA}
\end{figure*}

\begin{figure*}
    \centering
    \includegraphics[width=\linewidth]{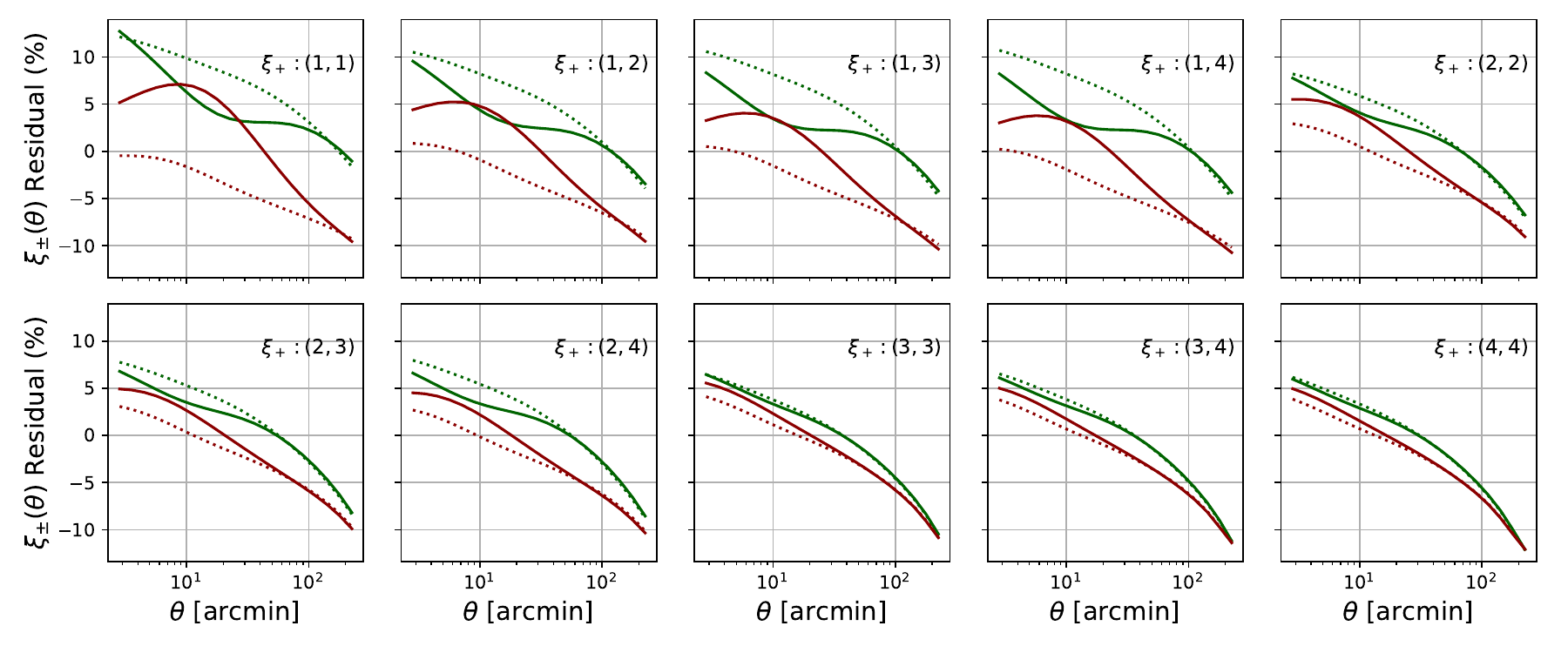}
    \includegraphics[width=\linewidth]{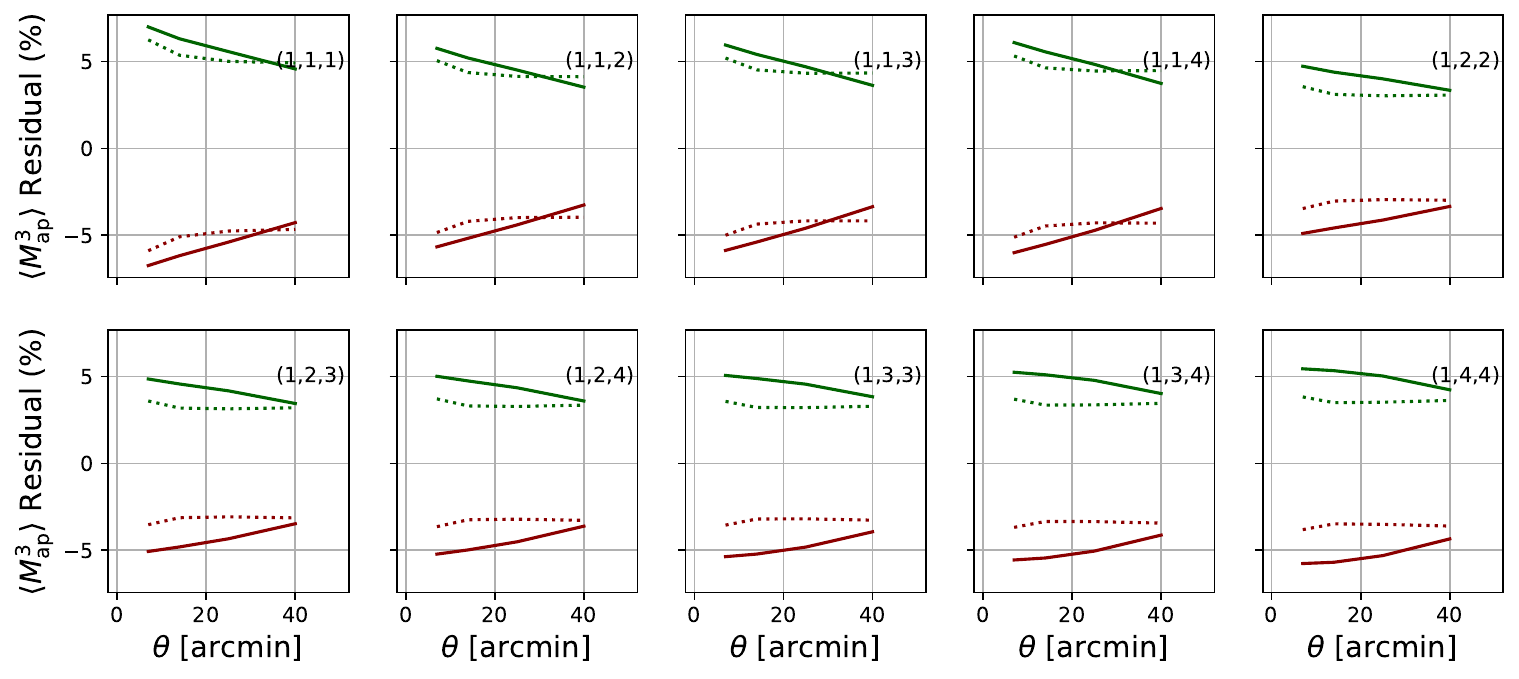}
    \caption{
    Residuals for $\xi_{+}$ (upper panels) and $\mapcube$ (lower panels) assuming different parameter sets and IA models. The solid lines correspond to TATT, while the dashed lines correspond to NLA. In green, we show the result for $A_1=-0.2$ and, in the TATT case, $A_2=0.4$. In red, we show the result for $A_1=0.2$ and, in the TATT case, $A_2=-0.4$. All other parameters are fixed to their values listed on Table~\ref{table:seccoparams}. We note that $\mapcube$ significantly helps to differentiate the signals from both parameter scenarios in the case of TATT. This effect is more significant for smaller scales, as the higher-order IA effects become less important for both $\xi_{+}$ and $\mapcube$ as the scale increases. The residuals are computed with respect to a model with no IA.}
    \label{fig:xip_and_map3}
\end{figure*}

We compute the total $\mapcube$ signal with varying values for the TATT intrinsic alignment parameters, in order to verify what values are required for our computation to significantly deviate from the NLA model. In Figure~\ref{varya2}, we leave the redshift evolution parameters $\alpha_1$ and $\alpha_2$ fixed at 1.0, and show the percent difference between the predictions with TATT and NLA for different values of $A_1$, $A_2$, and $b_{\text{ta}}$. We choose redshift combinations $(1,1,1)$ and $(1,3,4)$ to probe the effect of IA in low-redshift auto-correlations and cross-correlations between low and high redshfits. The auto-correlations at high redshifts are expected to carry smaller contributions from galaxy intrinsic alignments. 

Next, we add the velocity-shear contribution and show the $\mapcube$ percent difference when modeling IA with NLA and TATT+VS. For this test, we vary the ratio $C_t/C_1$, as both parameters are related in the Lagrangian evolution picture \cite{Schmitz_2018}. This ratio sets the relative strength of the velocity-shear effect with respect to the tidal alignment amplitude parameter. Our results are shown for redshift bin combinations $(1,1,1)$ and $(1,3,4)$ in Figure~\ref{varyct}.

In Figure~\ref{fig:paramcontributions}, we show the contribution of the different perturbative expansion terms to the total $\mapcube$ signal. We use the parameter values from Table~\ref{table:seccoparams} plus a velocity-shear contribution of $C_t=2.5C_1$. This choice is motivated by the Lagrangian linear alignment model, where the higher-order Lagrangian bias parameters are set to zero. In this scenario, we can use the co-evolution relations from Eq.~A10 of \citet{bakx} to set an approximate strength to the velocity shear effect. We set the galaxy bias to $b_s=1.0$ to arrive at this value. While the dominant contribution is from the first-order terms, the higher-order terms are non-negligible, with $C_2$ being more important than $C_{1\delta}$. For our chosen value of $C_t$, the velocity shear contribution has a similar order of magnitude to that of $C_2$, however with an opposite sign. We also investigate the composition of the higher-order contributions by splitting them into contributions from the different IA bispectra. We show our results in Figure~\ref{fig:bispeccontributions}, for which we maintain our fiducial set of model parameters, but do not include the velocity shear term. We verify that the contributions with two density fields and one shape field significantly surpass those with more than one shape field. Different permutations of the fields may lead to contributions to $\mapcube$ with different signs.
 
We also compute the total $\mapcube$ signal using parameter values from Table~\ref{table:seccoparams} to investigate a realistic scenario. In Figure \ref{fig:map3IA}, we show the percent strength of the $\mapcube$ intrinsic alignment contamination for NLA, TATT, and TATT+VS, all relative to a model with no IA. For TATT+VS, we maintain the value of $C_t=2.5C_1$. As in the case of the full 3PCF, the largest TATT contamination is found on combinations which involve the first redshift bin. While the impact of changing from NLA to TATT is smaller for $\mapcube$ than for $\Gamma_0$, it is still significant, and can be even larger when we add the velocity shear parameter.

In Figure \ref{fig:2pcfIA}, we show as a reference the ratio of $\xi_{\pm}$ computed with and without intrinsic alignment for the same set of cosmological and IA parameters. We note that moving from NLA to TATT can shift $\xi_{\pm}$ and $\mapcube$ in opposite directions. This can be seen clearly when comparing $\xi_+$ for bin combinations (1,i) and $\mapcube$ for bin combinations (1,i,j). For the former, the higher-order intrinsic alignment effects introduce a damping of the signal for most scales, while for the latter, they introduce a boost in the signal. We verify that this pattern also holds for different parameter combinations at the $1\sigma-2\sigma$ region of the IA parameter posteriors from DES Y3 [see \textit{TATT, no SR} contours on Fig. 15 of \citet{Amon.Weller.2021}].

The potential for joint 2PCF+3PCF analyses to break degeneracies between cosmological and nuisance parameters has been demonstrated in recent data studies \cite{Gomes.DES.data, Sugiyama.HSC.3PCF}. Our results further show that the TATT model introduces a significant modulation of third-order statistics that differs qualitatively from its effect on the 2PCF. Because intrinsic alignment signals scale differently with the lensing efficiency kernels in two-point versus three-point correlations, the joint analysis probes the alignment mechanism at different effective redshifts and projection weights. This multi-statistic approach may help self-calibrate the TATT parameters, potentially overcoming the limitations of simpler models like NLA, which may lack the complexity required to fully utilize the information gain from higher-order statistics. While a formal quantification of this self-calibration requires a full likelihood analysis, our findings provide the theoretical basis for such an improvement in future Stage-IV surveys.

We demonstrate the potential of degeneracy breaking by choosing a set of TATT parameters that is nearly degenerate for the 2PCF but clearly distinct for $\mapcube$. In Figure~\ref{fig:xip_and_map3}, we compare TATT predictions for $\xi_{+}$ and $\mapcube$ with $A_1, A_2 = (-0.2,0.4)$ and $A_1,A_2 = (0.2, -0.4)$. We also show the corresponding NLA predictions for $\xi_{+}$ and $\mapcube$ with $A_1=-0.2$ and $A_1=0.2$. For $\xi_{+}$, the NLA signal for the two scenarios is clearly distinct, while the TATT signal for the two parameter sets is similar. When we look at $\mapcube$, the opposite effect happens, and the difference between the signals with the TATT model is larger than that with NLA. This finding suggests that, besides the known degeneracy breaking potential of adding a higher-order statistic, a joint analysis of the 2PCF and the 3PCF has an additional advantage in constraining the parameters of the TATT model.


\section{Conclusion}

We build a model to compute the intrinsic alignment contribution to the cosmic shear three-point correlation function (3PCF) and the mass aperture skewness ($\mapcube$) under the tidal alignment and tidal torquing (TATT) formalism. We review effective field theory (EFT) approaches and connect them to the TATT model. We also include the velocity shear extension to the TATT model (TATT+VS). We compute the E and B mode contributions to the tree level bispectrum, and use this to calculate the natural components of the 3PCF through multipole decomposition of the bispectrum. We also calculate the $\mapcube$ from the 3PCF, providing a pathway to include TATT in joint $\xi_{\pm}$-$\mapcube$ analyses. We include comparisons to the simpler Nonlinear Alignment (NLA) model, which is a special case of TATT. 

We use numerical values from the best fit of the DES Y3 cosmic shear data to compare the impact of intrinsic alignments on the 3PCF and $\mapcube$ when using NLA, TATT, and TATT+VS. We find that the higher-order terms included in the TATT model are responsible for an additional structure to the 3PCF across triangle configurations. The impact on $\mapcube$ is smaller, due to the fact that it upweights equilateral contributions. 

We also explore the dependence of $\mapcube$ on different choices for the intrinsic alignment parameters. Our results show that differences between NLA and TATT/TATT+VS approach $10\%$ of the total $\mapcube$ signal for parameter values of the order of unity, which are reasonable choices due to the normalization implicit in the TATT parameter definitions \cite{Blazek.2019}.

Higher-order statistical analyses of weak lensing have recently been carried out for Stage III surveys, with results from $\mapcube$ yielding a factor of 2 improvement on the joint constraint between $\Omega_m$ and $S_8$
\cite{Burger.Martinet.2023,Gomes.DES.data,Sugiyama.HSC.3PCF}. 
These analyses however have relied on the more limited NLA modeling of intrinsic alignments. For Stage IV surveys, where we expect a higher signal-to-noise and tighter constraints, differences of the order of $10\%$ on the theoretical modeling become more significant. We show that in this context it will be essential to consider more complex intrinsic alignment models. Additionally, we show that $\xi_{\pm}$ and $\mapcube$ respond differently to combinations of $A_1$ and $A_2$, and that degeneracies between these two parameters can be broken when third-order statistics are added to the two-point functions. 

Our implementation of the TATT and TATT+VS contributions to the 3PCF and $\mapcube$ paves the way for higher-order statistical analyses with Stage III and IV data. We leave the application to data for future work.

\begin{acknowledgments}
RCHG and BJ are  partially supported by the US Department of Energy grant DE-SC0007901. JB is partially supported in this work by NSF awards AST2206563 and AST2442796 and DOE grant DE-SC0024787. Part of this work was supported by the NASA ROSES grant 22-ROMAN11-0011 via a JPL subaward. This publication is part of
the project “A rising tide: Galaxy intrinsic alignments
as a new probe of cosmology and galaxy evolution”
(with project number VI.Vidi.203.011) of the Talent programme Vidi which is (partly) financed by the Dutch
Research Council (NWO).
\end{acknowledgments}

\bibliography{refs}

\end{document}